\documentclass[aps,amsfonts,amsmath,prd,preprint,nofootinbib,11pt]{revtex4-1}
 
\usepackage{epsfig,bbm,cancel}
\usepackage{slashed,xcolor}
\usepackage[normalem]{ulem} 
\usepackage{amssymb}
\usepackage{amsmath}
\usepackage{amsthm}
\usepackage{mathtools}
\usepackage[english]{babel}
\usepackage[a4paper, total={7in, 10in}]{geometry}
\usepackage[skip=4pt plus1pt, indent=12pt]{parskip}

\begin{document}
\title{A Unified Approach To Find The Generalized Maxwell-Chern-Simons-Higgs BPS Vortices and Their Properties}

\author{Emir Syahreza Fadhilla$^{1,2}$}
\email{30221012@mahasiswa.itb.ac.id}
\author{Laurenzius Yudha Prasetyatama$^1$}
\email{yudhaprasetyatama@gmail.com}
\author{Bobby Eka Gunara$^{1,2}$}
\email{bobby@itb.ac.id}
\author{Ardian Nata Atmaja$^2$}
\email{ardi002@brin.go.id}

\affiliation{$^1$Theoretical High Energy Physics  Research Division, Institut Teknologi Bandung,
Jl. Ganesha 10 Bandung 40132, Indonesia.}
\affiliation{$^2$ Research Center for Quantum Physics, National Research And Innovation Agency (BRIN),
Kawasan PUSPIPTEK Serpong, Tangerang 15314, Indonesia.}

\begin{abstract}
In this work, we propose that all BPS vortex solutions within the generalized Maxwell-Chern-Simons-Higgs (MCSH) model can be found from a single system of equations. This set of equations is derived using the BPS Lagrangian method, which is a more robust generalization of Bogomol'nyi's trick. We show that the known spherically symmetric BPS vortices can be reproduced as certain limits of Bogomol'nyi's equations in the generalized MCSH Model. This provides us with a possible classification system using the auxiliary functions, \(X_i\)s, in the BPS Lagrangian. 
Furthermore, we also study the properties of each known vortex through the numerical approach where we found that all of the vortices behave similarly under variations of their free parameters and a system of well-separated MCSH vortices saturates the BPS bound.
\end{abstract}

\maketitle

\thispagestyle{empty}
\setcounter{page}{1}
\tableofcontents

\newpage

\section{Introduction}
The field theories in \(1+2\) dimensional spacetime have been actively studied recently, especially, as models for many condensed matter systems and dual theories in high energy physics. One of the most studied theories in \(1+2\) dimensional spacetime is the \(U(1)\) gauge-scalar theory where the scalar field, or Higgs field, is invariant under \(U(1)\) transformation. The \(U(1)\) gauge itself is known as the model for electromagnetism because the invariants can be interpreted as electromagnetic charge, and their respective gauge field dynamics resemble Maxwell's equations of electromagnetism. In the standard formulation, the dynamics of the gauge fields are encoded in the Lagrangian by the derivative terms that contain \(F_{\mu\nu}=\partial_\mu A_\nu-\partial_\nu A_\mu\), known as the electromagnetic tensor. There are only a few gauge-invariant terms that can be constructed in \(1+2\) dimensions. Still, it is well-known that there are two important gauge-invariant terms for this case, namely the Maxwell term that is proportional to the trace-norm of electromagnetic tensor, \(F_{\mu\nu} F^{\mu\nu}\), and the Chern-Simons term that is proportional to the Chern-Simons invariant which takes the form \(\epsilon^{\mu\nu\lambda}A_{\mu}F_{\nu\lambda}\) for Abelian gauge theories, and the non-linear terms that are proportional to \(F^4\) or higher are not dominant especially at low-energy regimes. In this context, the extended gauge theory, which contains both the Maxwell and Chern-Simons terms, is the massive gauge theory \cite{SCHONFELD1981157,Jackiw:1980kv,Deser:1981wh,Deser:1982vy,Deser:1984kw}. The gauge-scalar version of the generalized theory above has been proposed in \cite{Lee:1990eq,Lee:1991em} where the coupling with the Higgs field is introduced, and, more recently, it is shown that non-minimal couplings that are assumed to be dependent on the scalar field can be introduced which results in a new model called the generalized Maxwell-Chern-Simons-Higgs (MCSH) model \cite{bazeia2012generalized}. 

The simplest submodel in the \(U(1)\) gauge-scalar theory is the Maxwell-Higgs model where the gauge field couples minimally through the covariant derivatives and the dynamics of the gauge field only came from the Maxwell term. Suppose we chose the scalar potential term to be the polynomial potential. In that case, the Maxwell-Higgs model is equivalent to the Ginzburg-Landau theory of superconductivity where the Higgs field is physically interpreted as the order parameter such that the norm of the Higgs field is the number density of Cooper-pair condensate. The components \(F_{\mu\nu}\) are the electromagnetic fields in the two-dimensional superconductors \cite{Ginzburg:1950sr}. The Ginzburg-Landau theory has been proven as a good model for understanding the behaviour of type-II superconductors \cite{tinkham2004introduction}. Furthermore, an extended version of this theory with \(U(1)\times U(1)\) symmetry is proven to have unique cosmic string solutions \cite{Yang:2014hda,Han:2015yua}. A natural way to extend the effective field theory for these systems is to consider the generalized MCSH model, especially if we want to keep the existence of topological vortices as one of the main characteristics of superconductors. There exists further generalization of the gauged scalar model beyond the couplings of Maxwell term and Chern-Simons term that are already well-studied, for example, the Born-Infeld-Higgs model where the Lagrangian is proportional to \(\sqrt{\det(\eta+(F/b)}\), \(\eta\) is the Minkowski metric, and the \(K\)-generalized model where the norms of either \(F_{\mu\nu}\) or \(D_\mu\phi\) enters the Lagrangian as a function \(K(\|F\|)\) or \(K(\|D\phi\|)\). These models are also known to have vortex solutions \cite{Shiraishi:1990zi,Babichev:2007tn,Adam:2008rf,Bazeia:2010vb,Bazeia:2011bz,Casana:2015bea,Bazeia:2017gub,Bazeia:2018chj,Atmaja:2018ddi}. In this work we focus on the generalized MCSH model because the Lagrangian is guaranteed to contain the terms that are, at most, quadratic in first-derivatives of the fields with no higher power terms.

It is known that some of the non-perturbative solutions in the gauge-scalar theories are particle-like configurations that obey some extra conservation properties which came from the topology of the fields \cite{Vilenkin:2000jqa,Manton:2004tk}. These solutions are known as topological solitons which generally possess a useful property, namely, the topologically-supported stability, where solitons with non-zero topological charge cannot disperse into vacuum spontaneously. This stability property came from the BPS bound which states the lowest energy possible for a topological soliton, such that its energy cannot be less than the BPS energy \cite{Bogomolny:1975de}. An interesting class of topological solitons in \(1+2\) dimensional theory is the vortices which are the solitons that are characterized as zeros of the scalar fields with phase shifts linear to an integer if we trace the scalar fields in a closed loop surrounding the centre of the vortex \cite{Manton:2004tk}. 

The families of vortex solutions in the submodels of the MCSH model are well-known. One of the first vortex solutions known in the gauge-scalar theory is the Maxwell-Higgs (MH) vortex which is the topological soliton of the Ginzburg-Landau theory \cite{ABRIKOSOV1957199,Nielsen:1973cs,Taubes:1979tm} and the BPS solutions with non-minimal coupling have been found more recently \cite{Bazeia:1992cw,Bazeia:2011stq,Casana:2014qfa,Alonso-Izquierdo:2015tta,Bazeia:2016dmc,Bazeia:2018srv,Bazeia:2018hyv,Marques:2018xow,Lima:2020zoo,Bazeia:2020quv,Andrade:2020nsd,Andrade:2021uiu,Lima:2021gsl,AlonsoIzquierdo:2022zmw}
. The Chern-Simons-Higgs (CSH) vortices have also been studied as an alternative to the MH vortices with simpler field dynamics \cite{Jackiw:1990aw, Hong:1990yh} and their self-dual (BPS) solutions have been discussed in detail in \cite{Bazeia:2010wr,Bazeia:2017vzq}. For the more general cases where both MH and CS terms are non-zero and an additional real scalar field is introduced, the self-dual solutions were initially proposed in \cite{Lee:1990eq, Lee:1991em} and, further, the BPS solutions of the generalized case with non-minimal coupling constants have been found in \cite{bazeia2012generalized,Bazeia:2018hyv,Marques:2018xow,Andrade:2020nsd}. It is argued whether the real scalar field is necessary in order to find the BPS MCSH vortices, but a recent work \cite{Andrade:2021qkq} shows that there exists one specific BPS vortex in the generalized MCSH model without additional real scalar field. Thus, we expect that more types of BPS MCSH vortices exist even for models without additional real scalar fields.

It is believed that all of the BPS vortices mentioned above can only be reproduced by considering different limits of the coupling constants at the Lagrangian level. In this work, we proposed that by using the BPS Lagrangian method \cite{Atmaja:2015umo, Atmaja:2020iti,PrasetyaTama:2021elj, Atmaja:2023ipu} we can have a unified approach to find the BPS vortices on the level of Bogomol'nyi equations and some corresponding constraints. Following this method, the set of equations that need to be solved is provided in Section III. We show in Section IV that the BPS Lagrangian which contains, at most, linear terms in the derivative of the effective fields is sufficient to reproduce all known classes of BPS vortex equations within the generalized MCSH model. Such a unified approach is useful for understanding the family of non-interacting vortices in the general \(U(1)\) gauge-scalar theory and the corresponding classification can be used to easily predict the non-perturbative solutions that are yet to be found within the theory. Furthermore, our proposed method here mainly applies algebraic methods to find the dynamical equations of the fields, which leads to easily recognizable patterns in the sets of Bogomol'nyi equations and their relation to the conserved currents of the theory. Thus, our unified approach offers a potential application of the method in the pedagogical aspects of the vortices in gauge theory and BPS solitons in general.

\section{Generalized Maxwell-Chern-Simons-Higgs Model}\label{sec:MCSH}
The generalized Maxwell-Chern-Simons-Higgs model can be described with the Lagrangian density,
\begin{equation}\label{L-model}
	\mathcal{L} = -\frac{h(|\phi|)}{4}F_{\mu\nu}F^{\mu\nu} - \frac{\kappa}{4}\epsilon^{\mu\nu\rho}A_{\mu}F_{\nu\rho} + w(|\phi|)|D_{\mu}\phi|^{2} - V(|\phi|)
\end{equation}
where $F_{\mu\nu} = \partial_{\mu}A_{\nu} - \partial_{\nu}A_{\mu}$ is electromagnetic field strength tensor and $D_{\mu}\phi = \partial_{\mu}\phi + iA_{\mu}\phi$ is covariant derivative of the Higgs field. $h(|\phi|)$ denotes the generalized magnetic permeability and $w(|\phi|)|D_{\mu}\phi|^{2}$ is the generalized kinetic term of the Higgs field. Both $h(|\phi|)$ and $w(|\phi|)$ are assumed to be positive-definite functions. This model is endowed with potential $V(|\phi|)$ which the explicit form will be derived later. Moreover, the energy-momentum tensor corresponding with this Lagrangian can be written as
\begin{equation}\label{energy-momentum}
    T^{\mu\nu} = h(|\phi|)\left(F^{\mu\rho}F_{\rho}^{\;\nu} + \frac{1}{4}\eta^{\mu\nu}F_{\sigma\lambda}F^{\sigma\lambda}\right) + \eta^{\mu\nu}V + w(|\phi|)\left(2\mathcal{R}\left[\overline{D^{\mu}\phi}D^{\nu}\phi\right] - \eta^{\mu\nu}|D_{\lambda}\phi|^{2}\right)
\end{equation}
with $\mathcal{R[\cdot]}$ denotes the real part. It is straightforward to show that the Lagrangian \eqref{L-model} is invariant under \(U(1)\) gauge transformations
\begin{equation}
    \phi\rightarrow e^{i\psi}\phi,~~~A_\mu\rightarrow A_\mu-\partial_\mu\psi,
\end{equation}
by realizing that the transformation of the Chern-Simons term gives a boundary term and, hence, evaluates to zero by Gauss theorem. The Chern-Simons term also does not contribute to the energy-momentum tensor since the term does not depend on the metric which means that the term vanished under variation with respect to the metric. It is useful to introduce a characteristic energy scale \(\mathcal{E}\) such that the couplings and coordinates can be transformed into \(x^\mu\rightarrow x^\mu/\mathcal{E}\), \(h\rightarrow\mathcal{E}^3 h\), \(\kappa\rightarrow \mathcal{E}^4\kappa \), \(w\rightarrow \mathcal{E}w\) and the fields transform as \(A_\mu\rightarrow A_\mu/\mathcal{E}\), \(V\rightarrow \mathcal{E}^3 V\), where the scalar field is assumed to be dimensionless order parameter. Under such transformation \(F_{\mu\nu}\) become dimensionless, implying that we can transform the Lagrangian \eqref{L-model} by \(\mathcal{L}\rightarrow \mathcal{E}^3 \mathcal{L}\) and the energy-momentum tensor \eqref{energy-momentum} by \(T^{\mu\nu}\rightarrow \mathcal{E}^3 T^{\mu\nu}\) to become dimensionless quantity.

From this model, we can obtain the dynamics of the Higgs field via Euler-Lagrange equation, which gives
\begin{equation}\label{Higgs-dynamics}
	D_{\mu}\big(w(|\phi|)D^{\mu}\phi\big) + \frac{1}{4}\frac{\partial h}{\partial \overline{\phi}}F_{\mu\nu}F^{\mu\nu}-\frac{\partial w}{\partial \overline{\phi}}|D_{\mu}\phi|^{2} + \frac{\partial V}{\partial \overline{\phi}} = 0
\end{equation}
Similarly, the equation for the gauge field is
\begin{equation}\label{Gauge-dynamics}
	\partial_{\mu}(hF^{\mu\nu}) + J^{\nu} = \kappa F^{\nu}
\end{equation}
where $J^{\nu} = iw\big(\overline{(D^{\nu}\phi)}\phi - \overline{\phi}(D^{\nu}\phi)\big)$ and $F^{\nu} = (1/2)\epsilon^{\nu\alpha\beta}F_{\alpha\beta}$. Gauss's Law for the electric field in this model can be written as \(\partial_{j}(h\partial^{j}A_{t}) + 2wg^{2}A_{t} = \kappa F_{r\theta}.\)
From this relation, we can tell that the temporal gauge cannot be used since $A_{t} = 0$ will not solve the above equation. To deal with this condition, several attempts have been made before by modifying the model in order to increase the degree of freedom \cite{bazeia2012generalized,Andrade:2020sbl}.

We are interested in the radially symmetric static solution of the form,
\begin{equation}\label{ansatz}
	\phi(r,\theta) =  g(r)e^{in\theta};\qquad \textbf{A}(r) = -\frac{\hat{\theta}}{r}\big(a(r)-n\big);\qquad A_{t} = f(r)
\end{equation}
where $(r,\theta)$ are ordinary polar coordinates, and $n = \pm 1, \pm 2, ...$ is the winding number. Each field must obey the boundary condition
\begin{equation}\label{boundaryVal}
	g(0) = 0,\qquad a(0) = n,\qquad f(0) = {f_0}.
\end{equation}
For further calculation, we use the metric signature $(+,-,-)$. 

The Lagrangian in (\ref{L-model}) can also be written in terms of the defined ansatz. From which we obtain
\begin{equation}\label{L-effective}
	\mathcal{L} = \frac{h}{2}{f'}^{2} - \frac{h}{2}\left(\frac{a'}{er}\right)^{2} - \frac{\kappa f}{2}\frac{a'}{r} + \frac{\kappa(a-n)}{2r}{f'} - w\left({g'}^{2} + \frac{{a^2}{g^2}}{r^2}\right) + {f^2}{g^2}w - V(g)
\end{equation}
where the primed functions denote their derivative with respect to $r$. This effective lagrangian obeys spherical symmetry, as expected from the choice of ansatz, hence restricting ourselves to only spherically symmetric solutions of the MCSH model. From here, we can expect the energy-momentum tensor to obey the same symmetry. First, we consider the components of the energy-momentum tensor in spherical coordinates by substituting the ansatz \eqref{ansatz},
\begin{eqnarray}
T^{tt}&=&\frac{h}{2}{f'}^{2} + \frac{h}{2}\left(\frac{a'}{r}\right)^{2} + w\left({g'}^{2} + \frac{{a^2}{g^2}}{r^2}\right) + {f^2}{g^2}w + V(g)\\
T^{rr}&=&-\frac{h}{2}{f'}^{2} + \frac{h}{2}\left(\frac{a'}{r}\right)^{2} + w\left({g'}^{2} - \frac{{a^2}{g^2}}{r^2}\right) + {f^2}{g^2}w - V(g)\label{Trr}\\
T^{\theta\theta}&=&\frac{h}{2}{f'}^{2} + \frac{h}{2}\left(\frac{a'}{r}\right)^{2} - w\left({g'}^{2} - \frac{{a^2}{g^2}}{r^2}\right) + {f^2}{g^2}w - V(g)\label{Ttt}\\
T^{tr}&=&T^{rt}=T^{t\theta}=T^{\theta t}=T^{r\theta}=T^{\theta r}=0.
\end{eqnarray}
The \(tt\) component of the energy-momentum tensor is the most important quantity for our analysis since it is the energy density of the system of vortices. The rest of the components are used as extra constraints for the Bogomol'nyi equation by assuming that all spatial components of the energy-momentum tensor vanish which physically implies that the system is pressure-less. This pressure-less condition is a necessary condition for the stability of topological solitons \cite{Bazeia:2008tj,BAZEIA2009402,Fadhilla:2020rig,Atmaja:2020iti}.  

\section{The BPS Lagrangian Method}
This section aims to find the first-order field equations of the MCSH model. The BPS Lagrangian method is actually a more robust extension of the usual Bogomol'nyi's trick \cite{Atmaja:2015umo}. Similar to Bogomol'nyi's trick, the method implies the saturation of BPS bound,
\begin{equation}
    E\geq E_{BPS},~~~E_{BPS}=\int \rho_{BPS}~ d^3x,
\end{equation}
where \(\rho_{BPS}\) is the lower bound of the total static energy density, which coincides with the lower bound of Hamiltonian density, \(H\), in this case. We know that the Hamiltonian density is equivalent to the Lagrangian of the model for static cases, \(H=-\mathcal{L}\). Thus, the BPS limit, where the total static energy, \(E\), saturates the BPS bound, is achieved when \(\mathcal{L}=-\rho_{BPS}\). 
We can use this to define the BPS Lagrangian, \(\mathcal{L}_{BPS}\) that satisfies
\begin{equation}
    \mathcal{L}=-\sum_i |\partial_\mu\Psi^i-\mathcal{A}^i_\mu(\Psi,\partial_\mu\Psi,x^\mu)|^2+\mathcal{L}_{BPS},
\end{equation}
such that the Lagrangian is bounded from above by \(\mathcal{L}\leq \mathcal{L}_{BPS}\), where \(\Psi^i\)s are the dynamical fields and \(\mathcal{A}^i\)s are explicit vectors that are functions of the dynamical fields, their derivatives, and coordinates, which can be found from the Bogomol'nyi equations of each \(\Psi^i\). Now that we have the effective Lagrangian of the model, given in \eqref{L-effective}, we can identify each \(\Psi^i\) as \(g\), \(a\), and \(f\), and we can also assume that \(\mathcal{A}^i\) does not depend on the angular coordinate due to spherical symmetry. The earlier construction of the BPS Lagrangian method assumes that the quadratic term is sufficient to reproduce the second-order field equations from variational principle and the \(\mathcal{L}_{BPS}\) does not contribute to the field equation because it usually consists of boundary terms related to some conserved currents \cite{Atmaja:2015umo,Adam:2016ipc}. This assumption is necessary because, in order to have consistent first-order equations that are equivalent to their second-order counterpart, the Euler-Lagrange equations of \(\mathcal{L}\) must be the derivative of the first-order (Bogomol'nyi) equations. We can extend this assumption in a more general setting where \(\mathcal{L}_{BPS}\) does not consist of boundary terms and we can get rid of any contribution from \(\mathcal{L}_{BPS}\) to the field equations by solving it separately. Thus, the resulting system of equations is the set of Bogomol'nyi equations equipped with extra constraint equations from the Euler-Lagrange equations of \(\mathcal{L}_{BPS}\).

To employ the BPS Lagrangian method in the MCSH model, we begin by proposing first order differential equation of the involved fields as
\begin{align}
    &{g'} = {\mathcal{A}_1}(g,a,f,{a'},{f'},r)\\
    &\frac{a'}{r} = {\mathcal{A}_2}(g,a,f,{g'},{f'},r)\\
    &{f'} = {\mathcal{A}_3}(g,a,f,{g'},{a'},r)
\end{align}
The energy density of this model is the \(00\)-component of the energy-momentum tensor, from which we may write
\begin{equation}
    \begin{split}
        \rho = \frac{h}{2}\left({f'} - {\mathcal{A}_3}\right)^{2} + \frac{h}{2}\left(\frac{a'}{r} - \mathcal{A}_2\right)^{2} + w\left({g'} - {\mathcal{A}_1}\right)^{2} + h{A_3}{f'} + h{\mathcal{A}_2}\frac{a'}{r} + 2w{\mathcal{A}_1}{g'} \\- \frac{h}{2}{\mathcal{A}_3^2} - \frac{h}{2}{\mathcal{A}_2^2} - w{\mathcal{A}_1^2} + \frac{{a^2}{g^2}w}{r^2} + {g^2}{f^2}w + V
    \end{split}
\end{equation}
where the lower bound,
\begin{equation}
    \begin{split}
        \rho \geq h{\mathcal{A}_3}{f'} + h{\mathcal{A}_2}\frac{a'}{r} + 2w{\mathcal{A}_1}{g'} - \frac{h}{2}{\mathcal{A}_3^2} - \frac{h}{2}{\mathcal{A}_2^2} - w{\mathcal{A}_1^2} + \frac{{a^2}{g^2}w}{r^2} + {g^2}{f^2}w + V,
    \end{split}
\end{equation}
is saturated if all the quadratic terms are equal to zero. These zero quadratic terms are the first-order differential equations (Bogomol'nyi Equations) of \((g,a,f)\), and their solutions can be used to calculate the profile of the vortices.
This lower bound indicates that we may write the BPS Lagrangian as
\begin{equation}\label{General BPS Lagrangian}
    \begin{split}
        \mathcal{L}_{BPS} = -{X_0} - {X_1}{g'} - {X_2}{a'} - {X_3}{f'}  - {X_4}{g'}{a'} - {X_5}{g'}{f'} - {X_6}{a'}{f'} - {X_7}{g'}^{2} - {X_8}{a'}^{2} \\ - {X_9}{f'}^{2} - {X_{10}}{g'}{a'}{f'} - {X_{11}}{a'}^{2}{f'}^{2} - {X_{12}}{g'}^{2}{f'}^{2} - {X_{13}}{g'}^{2}{a'}^{2}
    \end{split}
\end{equation}
where all the constraint functions $X'$s $\equiv X(g,a,f,r)$ with the simplest one being the first four terms of the above BPS Lagrangian obtained for the $A'$s that are independent of the fields derivative.

In this work, we attempt to obtain the first-order equations with the simplest case. Let us simplify further the BPS Lagrangian as
\begin{equation}\label{L-BPS}
    \mathcal{L}_{BPS} = -{X_0}(g,a,f) - \frac{{X_1}(g,a,f)}{r}{g'} - \frac{{X_2}(g,a,f)}{r}{a'} - \frac{{X_3}(g,a,f)}{r}{f'}
\end{equation}
where the dependence of radial coordinate is assumed to be \(1/r\) such that the product of the BPS Lagrangian and the volume measure is free from the explicit radial coordinate. This assumption is useful to simplify the constraint equations coming from the Euler-Lagrange equations of \(\mathcal{L}_{BPS}\). Furthermore, since all of the \(X_i\)s do not explicitly depend on \(r\) in the Lagrangian level, we can separate each of the factors in front of \(r^m\) before any calculation using variational principle is done. In the BPS limit, effective Lagrangian and BPS Lagrangian must be equal, $\mathcal{L} - \mathcal{L}_{BPS} = 0.$
\begin{align}
	\begin{split}
		0=& \frac{h}{2}{f'}^{2} + \left(\frac{\kappa(a-n)}{2} + X_3\right)\frac{f'}{r} - \frac{h}{2}\left(\frac{a'}{r}\right)^{2} - \left(\frac{\kappa f}{2} - {X_2}\right)\frac{a'}{r} - w{g'}^{2} + \frac{X_1}{r}{g'}\\ &- \frac{w{g^2}{a^2}}{r^2}
		+{g^2}{f^2}w - V + X_0 .
	\end{split}
\end{align}
We shall consider the equation above as a quadratic equation of \(g',a',f'\) and a polynomial equation in \(r\). Thus,
by solving the above equation for each of the effective fields and assuming that each solution has zero discriminant, we obtain the Bogomol'nyi equations through an algebraic approach as follows
\begin{align}
	{g'}_{\pm} &= \frac{X_1 }{2wr},\label{BogEqg}\\
    \left(\frac{a'}{r}\right)_{\pm} &= \frac{2{X_2} - \kappa f}{2h},\label{BogEqa}\\
    {f'}_{\pm} &= \frac{-2{X_3} - \kappa(a-n) }{2rh},\label{BogEqf}
\end{align}
and the requirement that the discriminants are zero for each solution gives,
\begin{align}
	0 = \left(\frac{\kappa(a-n)}{2r} + \frac{X_3}{r}\right)^{2}- 2h\Bigg[\frac{X_1^2}{4w{r^2}} - \frac{w{g^2}{a^2}}{r^2} + \frac{(\kappa f - 2{X_2})^2}{8h} + {f^2}{g^2}w + {X_0} - V\Bigg]
\end{align}
This condition must be satisfied and this gives the equation that governs the constraint functions. Since the constraint functions do not depend on the coordinates explicitly, we shall argue that the coefficient of each term in the power of $r$ must vanish. From this we have,
\begin{align}
	&\label{const-eq1}{r^{-2}}: \frac{X_1^2}{4w}  - \frac{(\kappa(a-n) + 2{X_3})^2}{8h} - w{g^2}{a^2} = 0\\
	&\label{const-eq2}{r^0}: \frac{(\kappa f - 2{X_2})^2}{8h} + {f^2}{g^2}w + {X_0} - V = 0
\end{align}

On the other hand, the defined BPS Lagrangian (\ref{L-BPS}) must satisfy the Euler-Lagrange equations independently. Thus, we vary the Lagrangian with respect to each effective field to get
\begin{align}\label{constraint1}
	&\frac{d{X_1}}{dr} = r\frac{\partial X_0}{\partial g} + \frac{\partial X_1}{\partial g}{g'} + \frac{\partial X_2}{\partial g}{a'} + \frac{\partial X_3}{\partial g}{f'}\\ \label{constraint2}
	&\frac{d X_2}{dr} = r\frac{\partial X_0}{\partial a} + \frac{\partial X_1}{\partial a}{g'} + \frac{\partial X_2}{\partial a}{a'} + \frac{\partial X_3}{\partial a}{f'}\\ \label{constraint3}
 &\frac{d X_3}{dr} = r\frac{\partial X_0}{\partial f} + \frac{\partial X_1}{\partial f}{g'} + \frac{\partial X_2}{\partial f}{a'} +  \frac{\partial X_3}{\partial f}{f'}
\end{align}
Upon substitution of the BPS equations, we can rewrite the above results as below.
\begin{align}
	&\label{const-eq3}\frac{\partial X_0}{\partial g} = \frac{2{X_2} - \kappa f}{2h}\left(\frac{\partial X_1}{\partial a} - \frac{\partial X_2}{\partial g}\right),\\
	&\label{const-eq4}\frac{\partial X_0}{\partial a} = 0,\\
	&\label{const-eq5}\frac{\partial X_0}{\partial f} = \frac{2{X_2} - \kappa f}{2h}\left(\frac{\partial X_3}{\partial a} - \frac{\partial X_2}{\partial f}\right),\\
	&\label{const-eq6}\frac{2{X_3} + \kappa(a-n)}{2h}\left(\frac{\partial X_1}{\partial f} - \frac{\partial X_3}{\partial g}\right) = 0,\\
	&\label{const-eq7}\frac{X_1}{2w}\left(\frac{\partial X_3}{\partial g} - \frac{\partial X_1}{\partial f}\right) = 0,\\
	&\label{const-eq8}\frac{X_1}{2w}\left(\frac{\partial X_2}{\partial g} - \frac{\partial X_1}{\partial a}\right) = \frac{2{X_3} + \kappa(a-n)}{2h}\left(\frac{\partial X_2}{\partial f} - \frac{\partial X_3}{\partial a}\right).
\end{align}
All of the Bogomol'nyi equations from the MCSH model can be found by solving equations \eqref{const-eq1}, \eqref{const-eq2}, \eqref{constraint1}, \eqref{constraint2}, and \eqref{constraint3} for all \(X_0\), \(X_1\), \(X_2\), and \(X_3\). The resulting field equations could be either a Bogomol'nyi equation or a constraint equation that came from Euler-Lagrange equations of the \(\mathcal{L}_{BPS}\). 
\section{The Solutions}
In this section, we discuss some of the possible solutions of the equation system for \(X_0\), \(X_1\), \(X_2\), and \(X_3\) above. Here, we demonstrate how to reproduce the three well-known cases of the Chern-Simons-Higgs (CSH) vortex, Maxwell-Higgs (MH) vortex, and Maxwell-Chern-Simons-Higgs (MCSH) vortex from the set of Bogomol'nyi equations (\ref{BogEqg}-\ref{BogEqf}) and the constraint equations
(\ref{const-eq1}-\ref{const-eq8}) that are found from the BPS Lagrangian method on the generalized Maxwell-Chern-Simons-Higgs model.
\subsection{Chern-Simons-Higgs Vortex}
The first solution to be discussed is the Chern-Simons-Higgs (CSH) vortex with \(h=0\). This solution requires a different approach from the rest of the solutions because if we impose \(h=0\) then the Bogomol'nyi equations for \(a'\) and \(f'\) deduced above fail. This is due to the fact that the terms that are proportional to \((a')^2\) and \((f')^2\) are removed from \(\mathcal{L}\), hence we cannot complete their squares at the BPS limit. This fact implies that we are going to acquire the field equations for both \(a\) and \(f\) from the constraint equations.

Consider equations \eqref{const-eq1} and \eqref{const-eq2}. In order to achieve the Chern-Simons limit \(h=0\) we must impose
\begin{eqnarray}
    X_2=\frac{\kappa f}{2},~~~X_3=\frac{\kappa(n-a)}{2}.
\end{eqnarray}
Then, the two equations give us the solutions of both \(X_0\) and 
\(X_1\) as follows
\begin{eqnarray}
    X_0=V-f^2g^2w,~~~X_1=\pm 2wga.
\end{eqnarray}
From here, we are able to deduce that the Bogomol'nyi equations for \(g\) is given by
\begin{equation}
    g'=\pm \frac{ag}{r},\label{CSeqg}
\end{equation}
and the first constraint equation \eqref{constraint1} gives us the constraint on the potential term, namely
\begin{equation}
    \mp \frac{4f g^3 w^2}{\kappa} = - \frac{\partial V}{\partial g} +  f^2\frac{\partial (g^2 w)}{\partial g}.
\end{equation}
As we are going to see in the next section, the first order equation of \(g\) in for the Chern-Simons-Higgs vortex is exactly the same as the equation of \(g\) from the Maxwell-Higgs vortex. The field equations of the remaining fields, \(a\) and \(f\) can be deduced from the constraint equations \eqref{constraint2} and \eqref{constraint3}. Upon substitution of the \(X_i\)'s above, equation \eqref{constraint2} becomes
\begin{equation}
    f' = -\frac{2ag^2 w}{\kappa r}, \label{CSeqf}
\end{equation}
and equation \eqref{constraint3} becomes
\begin{equation}
    \frac{a'}{r} = -\frac{2 f  g^2 w}{\kappa}.\label{CSeqa}
\end{equation}
We should consider both equations \eqref{CSeqf} and \eqref{CSeqa} above as the field equations for \(a\) and \(f\). From the (\ref{CSeqf}), we may show that
       \( f' = \mp \frac{1}{\kappa} \frac{dG(g)}{dr},\)
where we have define a new function of \(g\), 
\begin{equation}\label{definG}
    \frac{dG}{dg} = 2 g w(g).
\end{equation} This result can be regarded as the self-dual equation for the temporal gauge field. Now that we have the expression for \(f\), the self-dual equation for the angular component of the gauge field can be obtained by substituting \(f\) into eq. (\ref{CSeqa}) and
the governing equation for the potential can be recast into \(\frac{\partial V}{\partial g} = \frac{1}{\kappa^2}\left(G^2 \frac{\partial(g^2 w)}{\partial g} + 4g^3 w^2 G\right)\). In a conclusion, we have the expression for \(f\) and the first-order equation of \(a\),
\begin{eqnarray}
   \label{CSeqfsimplest} f &=& \mp \frac{1}{\kappa} G(g),\\
     \label{CSeqasimplest} \frac{a'}{r} &=& \pm \frac{2 g^2 w}{\kappa^2} G(g),
\end{eqnarray}
and the constraint equation for the potential is simplified by introducing \(G\),
\begin{equation}
    \begin{split}
        \frac{\partial V}{\partial g}
        = \frac{\partial}{\partial g}\left(\frac{1}{\kappa^2} G^2 g^2 w\right)~~~\Rightarrow V(g) = \frac{1}{\kappa^2} G^2 g^2 w.\label{PotConstCS}
    \end{split}
\end{equation}
The system of equations for the effective fields \eqref{CSeqg}, \eqref{CSeqasimplest}, \eqref{CSeqfsimplest} and the constraints \eqref{definG}, \eqref{PotConstCS} are the first-order field equations for the Chern-Simons-Higgs vortex which have been found from a different scheme in \cite{Bazeia:2017nas}. We can see that the remaining freedom is in the choice of either the potential or the coupling constant and fixing one of them will fix the other. Furthermore, the system is effectively a system of coupled differential equations of only \(a\) and \(g\) since \(f\) depends directly on \(g\). Thus, in order to find a specific vortex solution we need to choose a form of potential and then proceed to solve the first-order equations of \(a\) and \(g\). An example of this is provided in Section \ref{Sec:NumCal}.

\subsection{Generalized Maxwell-Higgs Vortex}
The generalized Maxwell-Higgs vortices can be found by setting \(\kappa=0\) in both Bogomol'nyi equations and constraint equations. In contrast with the previous case of CSH vortex, the square of the first derivatives of \(a\) and \(f\) appear in the effective Lagrangian which leads to well-behaving differential equations of all \(X_i\)s. It is known that the vortices within this model do not emit any electric fields and we are going to show that such property can be derived from the BPS Lagrangian method below.

From (\ref{const-eq4}), we have $X_0 \equiv {X_0}(g,f)$ from which (\ref{const-eq2}) implies  $X_2 \equiv {X_2}(g,f)$. It also tells us that, from (\ref{const-eq3}), we can write $X_1$ as
\begin{equation}\label{X1-Y}
	X_1 = a{Y_1}(g,f) + {Y_2}(g,f)
\end{equation}
Similarly, from (\ref{const-eq5}) we can write $X_3$ as
\begin{equation}\label{X3-Z}
	X_3 = a{Z_1}(g,f) + {Z_2}(g,f)
\end{equation}
for some unknown functions, ${Y_1}, Y_2, Z_1$ and $Z_2$ such that we have the solutions for \(Y_2\) and \(Z_2\), namely
\begin{equation}
    Y_2=0,~~~Z_2=0,
\end{equation}
and two equations to be solved for \(Y_1\) and \(Z_1\), namely
\begin{eqnarray}
    0&=& g^2 w - \frac{Y_1^2}{4w} + \frac{Z_1^2}{2h} ,\label{eq:ConstraintEqM-1-a2}\\
    0&=& f \left(\frac{\partial}{\partial g}\left(\frac{Z_1}{f}\right) - \frac{2g^2 w^2 Z_1}{f Y_1}\frac{\partial}{\partial f}\left(\frac{Z_1}{g^2wh}\right)\right) .\label{MHZ1}
\end{eqnarray}
Equation \eqref{MHZ1} implies two possible cases. The first one is \(f=0\) and the second one is when we choose to take the factor inside the parenthesis to be zero. The Bogomol'nyi equations from the second case are not consistent with the second-order equations from Euler-Lagrange equations of \(\mathcal{L}\) (see. Appendix \ref{sec:SecondOrderEq}). Thus, we omit the possibility of the second case and focus ourselves on the first case of \(f=0\). Equation \eqref{MHZ1}, generally, possesses a large family of solutions. Thus, we need to constrain our solutions further by choosing only physically relevant solutions. This can be done by considering extra constrain that came from the energy-momentum tensor. Since the vortex solution needs to be stable, then, the angular components of the energy-momentum tensor must be null, i.e. there is no angular pressure for spherically symmetric vortices. This gives us
 \(X_0=0\).

If \(X_0=f=0\)  then from the Bogomol'nyi equation \eqref{BogEqf} we know that \(Z_1=0\), which implies that \(Y_1=\pm2gw\). From \eqref{X1-Y}, \eqref{X3-Z}, and \eqref{const-eq2}, we have the expression of the \(X_i\)s for the MH vortex, namely
\begin{equation}
    X_1=\pm 2agw,~~~X_2=\pm\sqrt{2hV},~~~X_3=0,
\end{equation}
implying that the Bogomol'nyi equations for this vortex are given by
\begin{subequations}
\begin{align}
    \label{BogEqMHg}{g'} &= \pm \frac{a g}{r},\\
    \label{BogEqMHa}\frac{a'}{ r} &= \mp \sqrt{\frac{2 V(g) }{h(g)}} .
\end{align}
\end{subequations}
Furthermore, equation \eqref{const-eq8} implies that \(X_2 (g) = \pm \int 2 g w(g) \; dg\), this can be used to specify the form of potential,
\begin{equation}\label{ConstMHPot}
    V(g) = \frac{2}{h(g)}\left(\int g w(g)\; dg\right)^2.
\end{equation}
One might notice that, in contrast with the CS vortex in the previous section where we have only a single degree of freedom, here we have two functions that are arbitrary, namely, the potential and the coupling \(h\). The resulting system of equations, \eqref{BogEqMHg}, \eqref{BogEqMHa}, and \eqref{ConstMHPot}, also resembles the standard MH vortex equations where \(h\) is a constant. As such, the correspondence to the Ginzburg-Landau theory of superconductivity is straightforward.

As mentioned in the earlier subsection of the CSH vortex, we can see that the Bogomol'nyi equation of \(g\) for the MH vortex, equation \eqref{BogEqMHg}, is exactly the same as the one we found in the CSH vortex, \eqref{BogEqg}, hence we should expect that the profile of the scalar field should be similar for both vortices. An example of the MH vortex profile is provided in Section \ref{Sec:NumCal}.

\subsection{Generalized Maxwell-Chern-Simons-Higgs Vortex}
In this section, we are going to derive the most general cases within this model where both the coupling constants \(\kappa\) and \(h\) take non-zero values. Such cases admit a wide family of Bogomol'nyi equations and one particular set of solutions that do not have the CSH and MH limit has been found in \cite{Andrade:2021qkq}. 

Let us proceed by showing the master equations that characterize the family of Bogomol'nyi equations in this case. As previously, equation \eqref{const-eq4} implies $X_0 \equiv {X_0}(g,f)$ which further, by equation \eqref{const-eq2},  imply $X_2 \equiv {X_2}(g,f)$ and furthermore from equations \eqref{const-eq5} and \eqref{const-eq2} we obtain 
\begin{equation}
    X_2=\frac{f \kappa }{2}+\frac{4 f g^2 h(g) w(g)}{\kappa -2Z_1}.
\end{equation}
As discussed in the previous section we will use expressions \eqref{X1-Y} and \eqref{X3-Z} for $X_1$ and $X_2$ respectively. Solving equation \eqref{const-eq8} for each power of $a(r)$, we extract two equations:
\begin{subequations}
\begin{eqnarray}
 &&  \frac{(\kappa  n-2 Z_2)}{h(g)}\left({\partial X_2 \over \partial f}-Z_1\right)+\frac{Y_2}{w(g)}\left({\partial X_2 \over \partial g}-Y_1\right) =0,\label{const-eq8-a0}\\
&& \frac{ (2 Z_1+\kappa )}{h(g)}\left(Z_1-{\partial X_2 \over \partial f}\right)+\frac{Y_1 }{w(g)}\left({\partial X_2 \over \partial g}-Y_1\right)=0.\label{const-eq8-a1}
\end{eqnarray}    
\end{subequations}
Using a linear combination of those equations, we can write the following equations
\begin{subequations}
    \begin{eqnarray}
      0&=&(Y_1 (\kappa  n-2 Z_2)+Y_2 (2 Z_1+\kappa ))  \left(8 g^2 h w \left(2 f {\partial Z_1 \over \partial f}-2Z_1+\kappa \right)+(\kappa -2 Z_1)^3\right),\\
      0&=&(Y_1 (\kappa  n-2 Z_2)+Y_2 (2 Z_1+\kappa )) \left(8 f g^2  h w {\partial Z_1 \over \partial g}-(\kappa -2 Z_1)\left(  Y_1\left(\kappa-2Z_1\right)-4 f  {\partial\left(g^2 w h\right)\over\partial g}\right)\right),    \end{eqnarray}
\end{subequations}
with a condition $Z_1\neq {\kappa\over 2}$.
These two equations can be reduced to one equation,
\begin{equation}
    Y_1 (\kappa  n-2 Z_2)+Y_2 (2 Z_1+\kappa )=0,
\end{equation}
and all together with the algebraic equations in \eqref{const-eq1} give us solutions
\begin{equation}
    Y_2=0,~~~Z_2=\frac{n\kappa}{2},~~~Y_1=\pm \sqrt{{1\over 2}(2Z_1+\kappa )^2+4 g^2 h w}\sqrt{w\over h}.
\end{equation}
However, these solutions only trivially satisfy equation \eqref{const-eq8-a0}, but not equation \eqref{const-eq8-a1}. Substituting these solutions into equations \eqref{const-eq6} and \eqref{const-eq7} and by demanding the resulting equations to be independent of the sign ($\pm$) we can conclude that $Z_1$ must be a constant, $Z_1=cz_1$ with $cz_1$ is a real constant, and hence $Y_1\equiv Y_1(g)$. 
Doing similarly to equation \eqref{const-eq8-a1}, with $Z_1$ constant, implies two equations
\begin{subequations}
    \begin{eqnarray}
        \frac{(2 cz_1+\kappa )^2}{4 h}+2 g^2 w&=&-\frac{(2 cz_1+\kappa ) }{4 h (\kappa -2 cz_1)}\left(8 g^2 h w +(\kappa -2cz_1)^2\right),\label{const-eq8-1}\\
        {\partial\left(g^2 w(g) h(g)\right)\over\partial g} &=&0.\label{const-eq8-2}
    \end{eqnarray}
\end{subequations}
Solution to \eqref{const-eq8-2} is $w={c_w\over g^2 h}$, with $c_w>0$ is a real constant, which after substituting  into  equation \eqref{const-eq8-1} yields $c_w={1\over 8}(4cz_1^2-\kappa^2)$ such that we obtain
\begin{equation}
    w(g) = \frac{4{cz_1^2} - {\kappa^2}}{8{g^2}h(g)},
\end{equation}
where $|cz_1| > |\frac{\kappa}{2}|$. The remaining constraint equation that needs to be solved is equation \eqref{const-eq3}.  Similarly by demanding equation \eqref{const-eq3} to be independent of the sign $(\pm)$, we obtain two equations:
\begin{subequations}
    \begin{eqnarray}
        -\frac{cz_1 f^2 (2 cz_1 +\kappa ) h'(g)}{2 h(g)^2}-V'(g)=0,\\
        \sqrt{cz_1} f (2 cz_1 +\kappa )^{3/2} \sqrt{4 cz_1^2-\kappa ^2}=0.
    \end{eqnarray}
\end{subequations}
Unfortunately, there is no solution to these equations since they require $|cz_1|=|{\kappa\over 2}|$ which leads to $w=0$. The only possible solution to equation \eqref{const-eq3} is if $f$ depends on $g$ , $f\equiv f(g)$ .
We can see the indication $f\equiv f(g)$ from the BPS equations which can be written as
\begin{align}
    &{g'} = \pm \sqrt{\frac{4{cz_1}}{2{cz_1}-\kappa}}\frac{ag}{r},\\
    &\frac{a'}{r} = -\frac{(2{cz_1}+\kappa)f}{2h},\\
    &\label{BPS-f}{f'} = -\frac{(2{cz_1}+\kappa)a}{2rh},
\end{align}
Taking $f'/g'$, we find that $f(g)$ is given by
\begin{equation}
    f=\mp\frac{(2 cz_1 +\kappa )\sqrt{(2 cz_1 -\kappa )}}{4 \sqrt{cz_1}}\int \frac{1}{g~ h(g)}dg.\label{f(g)}
\end{equation}
Substituting this into equation \eqref{const-eq3} gives us the solution for potential $V$ in terms of double integrals over $g$ which is a rather unusual form of potential. We may restrict further by requiring solutions to be physically stable by imposing the pressureless condition such that the radial and angular components of the energy-momentum tensor, equations \eqref{Trr} and \eqref{Ttt} respectively, to be zero, $T^{rr}=T^{\theta\theta}=0$. This condition implies $X_0=0$ and so, from equation \eqref{const-eq2} ,
\begin{equation}
    f=s_f \sqrt{2 V(g)~h(g) \over cz_1(2cz_1+\kappa)},
\end{equation}
where $s_f$ can be either $+1$ or $-1$. Comparing this with equation \eqref{f(g)} gives us potential
\begin{equation}
    V(g) = \frac{(2{cz_1}+\kappa)^3 (2{cz_1}-\kappa)}{32~ h(g)}\left(\int \frac{d g}{g~ h(g)}\right)^{2}.
\end{equation}
Surprisingly this potential, together with $f(g)$, solves the remaining constraint equation \eqref{const-eq3}.
This resulting system of equations is equivalent to the one derived in \cite{Andrade:2021qkq}. In order to show this, let us introduce a redefinition of $\kappa \rightarrow -\kappa$ and a new constant \(l\), satisfying
\begin{equation}
    {cz_1} = \frac{\kappa}{2}(2l+1),
\end{equation}
where $l>0$ and $l<-1$.
Substituting these new constants gives us
\begin{align}
    &{g'} = {s_g} \sqrt{\frac{2l +1}{l+1}}\frac{a g}{r}\\
    &\frac{a'}{r} = -{s_f} \sqrt{\frac{2lV}{(2l+1)h}}\\
    &f = {s_f} \frac{1}{\kappa}\sqrt{\frac{2hV}{l(2l+1)}}
\end{align}
with the potential constrained by \(h\), 
\begin{equation}
    V(g) = \frac{{\kappa^4}{l^3}(l+1)}{2h}\left(\int \frac{d g}{g h(g)}\right)^2.
\end{equation}
Similar to the CSH vortex, the only arbitrary function left to be fixed is $h$ which determine the form of potential of the model. Here we have derived rigorously BPS vortices in \cite{Andrade:2021qkq}.

\section{Numerical Calculations}\label{Sec:NumCal}
In this section, we provide some examples of the solutions with certain types of potential terms. Firstly, let us summarize the field equations that have been reproduced through the BPS Lagrangian method in previous sections. The first-order equations and the corresponding constraints for each vortex are provided in Table \ref{tab:equations}.
\begin{table}[h]
    \centering
    \begin{tabular}{|c||c|c|c|}
    \hline
         &CSH&MH&MCSH  \\
         \hline
         \(g\)&\(g'=\pm\frac{ag}{r}\)&\(g'=\pm\frac{ag}{r}\)&\(g'=\pm \sqrt{\frac{2l +1}{l+1}}\frac{a g}{r}\)\\
         \hline
         \(a\)&\(\frac{a'}{r}=\mp \frac{2}{\kappa}\sqrt{Vg^2w}\)&\(\frac{a'}{r}=\mp\sqrt{\frac{2V}{h}}\)&\(\frac{a'}{r}=\mp\frac{2}{\kappa}\sqrt{\frac{Vg^2w}{(2l+1)(l+1)}}\)\\
         \hline
         \(f\)&\(f=\pm\sqrt{\frac{V}{g^2w}}\)&0&\(f=\pm\sqrt{\frac{(l+1)V}{(2l+1)g^2w}}\)\\
         \hline
         \(V\)&~~\(V=\frac{4g^2w}{\kappa^2}\left(\int gw ~dg\right)^2\)~~&~~\(V=\frac{2}{h}\left(\int gw~dg\right)^2\)~~&~~\(V=\frac{2lg^2w}{(l+1)\kappa^2}\left(\int gw~dg\right)^2\)~~\\
         \hline
         \(h\)&0&arbitrary&depends on \(V\)\\
         \hline
         \(w\)&depends on \(V\)&depends on \(V\)&depends on \(V\)\\
         \hline
         Free Parameters&\(\kappa\)&\(h(g)\)&\(\kappa\) and \(l\)\\
         \hline
    \end{tabular}
    \caption{First-order field equations from the Bogomol'nyi and constraint equations, and the conditions for the couplings \(h\) and \(w\) corresponding to the potential for every model.}
    \label{tab:equations}
\end{table}
We would like to remark that the CSH and MCSH vortices have the same relation between \(w\) and \(V\), up to a constant that depends on \(l\). As such, for the sake of simplicity, we shall fix the \(h\) coupling of the MH vortex to satisfy
\begin{equation}\label{relatehwMH}
    h_{MH}=\frac{m}{g^2 w_{MH}},
\end{equation}
with \(m\in \mathbb{R}^+\) is an arbitrary constant, such that all of the potentials for every vortex solution satisfy the same relation for \(V\) and \(w\). Furthermore, from this point, we choose to work with the positive-signed equation for the Bogomol'nyi equations of \(g\) and the sign of the rest of the equations follows. The other possible sign has the same physical properties with the only difference being that the sign of their topological charge becomes negative, hence we can choose only one of them for numerical calculations without any loss of generality.

For the numerical calculations, it is worth noting that the boundary values given in \eqref{boundaryVal} for \(a\) and \(g\) are not well-defined at \(r=0\). As such, we need to translate the conditions at \(r=0\) into conditions for large \(r\) which must be satisfied by the functions asymptotically. In general, we want these functions to be finite at large \(r\). Thus, the strategy to solve the equations is by using the shooting method, with shooting parameters introduced at \(r=\epsilon\) very close to the origin \(r=0\) such that the boundary values for \(g\) and \(a\) at \(\epsilon\) are approximately
\begin{equation}
    g(\varepsilon)=\varepsilon g'(0)+O(\varepsilon^2),~~~a(\varepsilon)=n+\varepsilon a'(0)+O(\varepsilon^2).
\end{equation}
Since we know that \(a'(0)=0\) then we only have one shooting parameter which came from \(g\), namely
\begin{equation}
    \mathcal{P}=\lim_{r\rightarrow 0} \frac{g}{r},
\end{equation}
where we have to find the value of \(\mathcal{P}\) such that both \(g\) and \(a\) are, approximately, finite at large \(r\). We are going to assume that the vacuum expectation value of \(|\phi|\) is one and the magnetic field goes to zero far from the origin, hence the functions are asymptotic to 
\begin{equation}
    \lim_{r\rightarrow \infty}g=1,~~~\lim_{r\rightarrow \infty}a=0.
\end{equation}

There are four important physical quantities that need to be solved for static vortex solutions, namely
\begin{itemize}
    \item the profile of norm-squared scalar field, \(|\phi|^2=g^2\),
    \item the distribution of magnetic field, \(B=-\frac{a'}{r}\)
    \item the distribution of radial electric field, \(\textbf{E}=-\frac{df}{dr}\hat{\textbf{r}}\),
    \item and the energy density of the system, \(\rho\).
\end{itemize}
The norm-squared scalar field should be interpreted as the density of the condensate, the magnetic field \(B\) is the three-dimensional magnetic vector perpendicular to the two-dimensional system, and the radial electric field \(\textbf{E}\) is the three-dimensional electric vector field projected on the two-dimensional system that is generated by the charged vortex. The two electromagnetic fields contribute to the total static energy density and the electric field mediates the interaction between two or more charged vortices. To study the interaction between two vortices, we can calculate the total energy of two well-separated vortices with separation \(d\) and then proceed by varying the separation in order to see whether the energy is increasing or decreasing. As such, any increase or decrease in the total energy is attributed to the interaction energy between the two vortices.

\begin{figure}
    \centering
    \includegraphics[width=0.43\linewidth]{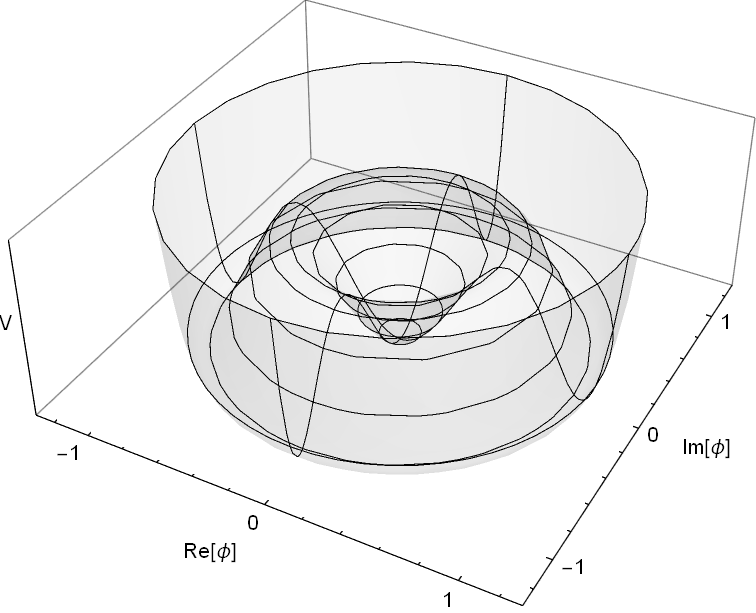}
    \includegraphics[width=0.43\linewidth]{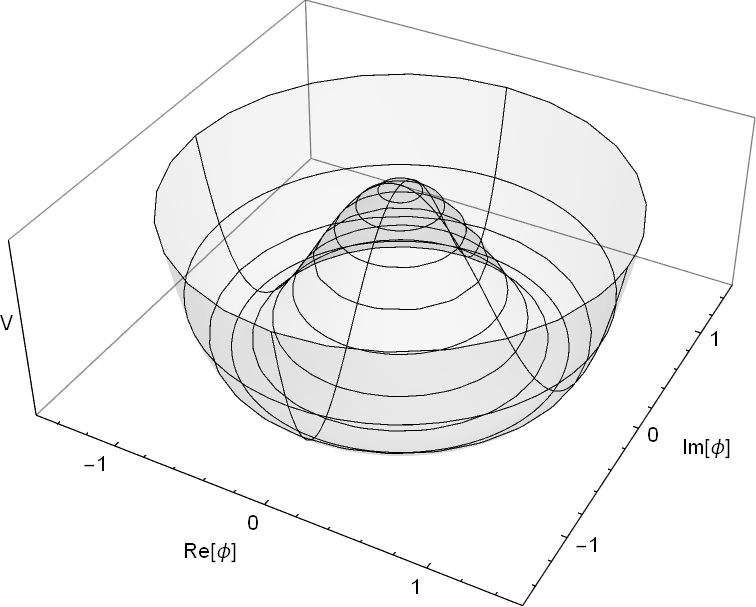}
    \caption{The profile of sextic potential (left) and quartic potential (right) as functions of the Higgs field, \(\phi\).}
    \label{fig:potential}
\end{figure}

We are going to consider two types of coupling \(w\), namely
\begin{itemize}
    \item \(w=1\) with potential \(V\propto g^2(1-g^2)^2\),
    \item \(w\propto\frac{(1-g^2)^2}{g^2}\left(\frac{3g^2(4-g^2)-12\ln{g}+c} {4}\right)^{-2/3}\) with potential \(V= (1-g^2)^2\).
\end{itemize}
The first type is called the sextic potential and the second type is called the quartic potential, which is also well-known as the Mexican hat potential. The profile for each of the potentials considered above is given in Fig. \ref{fig:potential}.
The sextic potential has vacuum expectation values at \(g=0\) and \(g=1\) but the quartic potential has only one stable vacuum, namely \(g=1\). Because the quartic potential has only one stable vacuum, there is only one possible solution for \(g\) which interpolates from the centre value \(g(0)=0\) to the boundary value \(g=1\) but this is certainly not the case for the sextic potential where \(g=0\) is also a stable vacuum, resulting in a non-topological solution. Thus, it is interesting to compare the numerical solution of the fields from these two different potential.
\subsection{Single Vortex Solutions}
\subsubsection{CSH Vortex}
\begin{figure}[h]
    \centering
    \includegraphics[width=0.32\textwidth]{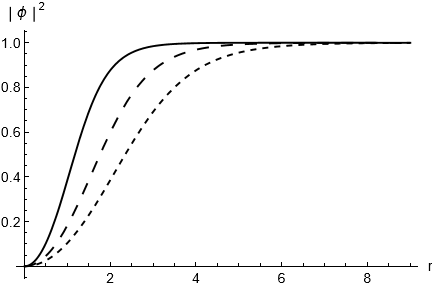}
    \includegraphics[width=0.32\textwidth]{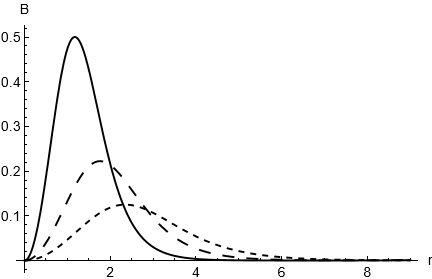}
    \includegraphics[width=0.32\textwidth]{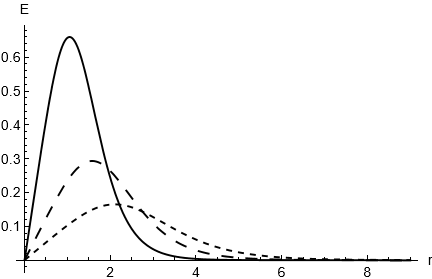}
    \caption{The profile electromagnetic fields and condensate density of CSH Vortices with sextic potential and the variation of \(\kappa=1\) (line), \(\kappa=1.5\) (dashed), and \(\kappa=2\) (dots).}
    \label{fig:CSH-single}
\end{figure}

\begin{figure}[h]
    \centering
    \includegraphics[width=0.32\textwidth]{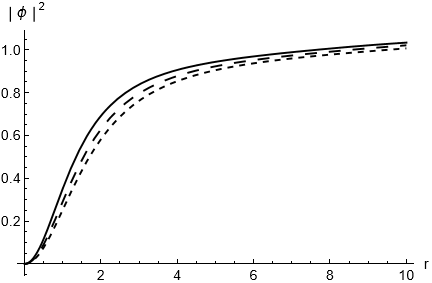}
    \includegraphics[width=0.32\textwidth]{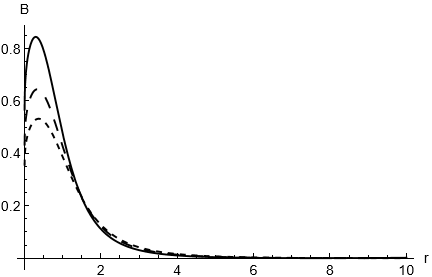}
    \includegraphics[width=0.32\textwidth]{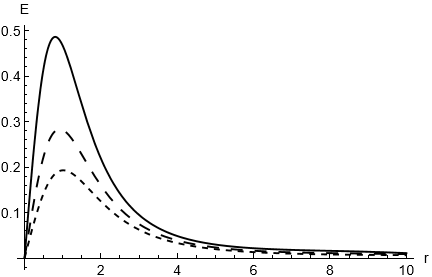}
    \caption{The profile electromagnetic fields and condensate density of CSH Vortices with quartic potential and the variation of \(\kappa=1\) (line), \(\kappa=1.5\) (dashed), and \(\kappa=2\) (dots).}
    \label{fig:CSH-single-quartic}
\end{figure}

The numerical solutions of the electromagnetic fields and the norm-squared scalar field of a single CSH vortex are shown in Figure \ref{fig:CSH-single} and Figure \ref{fig:CSH-single-quartic}. Here the only free parameter for variation is the Chern-Simons coupling constant, \(\kappa\), and we choose three ascending values for \(\kappa\), namely, \(1,~1.5,~\)and \(2\). 

For the first case with sextic potential, shown in the first figure in Fig. \ref{fig:CSH-single}, we can see that the norm-squared scalar field, which can be physically interpreted as the density of the condensate, is monotonically increasing with respect to \(r\) and the same feature can also be seen in Fig. \ref{fig:CSH-single-quartic}. Such behaviour of condensate is expected since the angular momentum of charged particles is, classically, higher for regions with strong magnetic fields. Thus, the region near the centre of the vortex is energetically less favourable, which explains why the profile is increasing for large \(r\). One might argue that this result came from the choice of the boundary conditions for \(g\), assumed in \eqref{boundaryVal}, but this is the only possible choice because the only possible boundary values of \(\phi\) are \(0\) and \(1\) which are the vacuum expectation values from the sextic potential. If we choose the centre value of \(g\) to be \(1\) then the vortex becomes non-topological and the stability is not guaranteed by the BPS bound. In fact, we can also assume that \(g(0)=1\) and \(g(r\rightarrow\infty)=0\) which represents the BPS solution with negative-signed Bogomol'nyi equations but it is obviously not a model for a physical two-dimensional material in a superconducting state, hence, we are going to only consider \(g(0)=0\) case in this work. 

The profile for the radial electric field and the magnetic field is similar and they get more spread in space as \(\kappa\) increases. This implies that the vortices has more interaction range for higher \(\kappa\). The peak of the magnetic and electric fields are also pushed away from the origin for higher values of \(\kappa\). Furthermore, the peak of the magnetic field is located near the steepest increase of the condensate density, as expected from the classical explanation from the previous paragraph.

An interesting observation is that the profile of vortices with quartic potential is less sensitive to the variations of coupling \(\kappa\) compared to the sextic potential cases. Furthermore, the profile of the electric and magnetic fields are asymptotic to a single function that does not depend on \(\kappa\). This property implies that we can define a characteristic size of the vortex that is independent of \(\kappa\). The peaks of the magnetic and electric fields are also located at different locations for quartic potential cases with the peak of magnetic field profile located closer to the origin. This leads to stronger electric and magnetic fields for the vortices with quartic potential cases compared to the sextic potential cases.

\subsubsection{generalized MH Vortex}
\begin{figure}[h]
    \centering
    \includegraphics[width=0.32\textwidth]{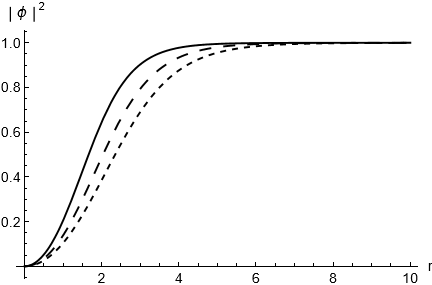}
    \includegraphics[width=0.32\textwidth]{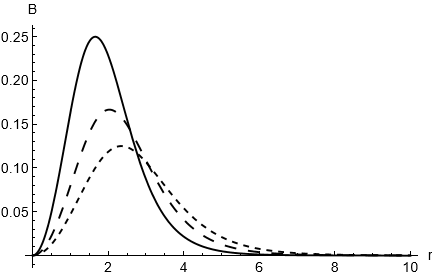}
    \caption{The profile magnetic fields and condensate density of MH Vortices with sextic potential and the variation of \(m=1\) (line), \(m=1.5\) (dashed), and \(m=2\) (dots).}
    \label{fig:MH-single}
\end{figure}
\begin{figure}[h]
    \centering
    \includegraphics[width=0.32\textwidth]{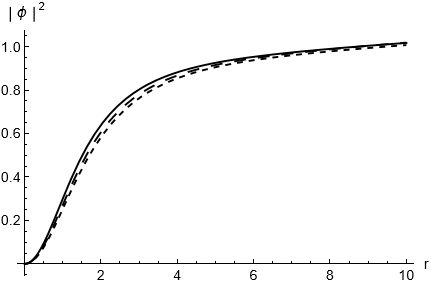}
    \includegraphics[width=0.32\textwidth]{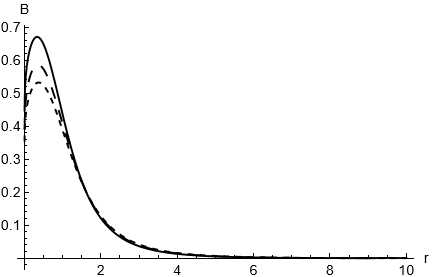}
    \caption{The profile magnetic fields and condensate density of MH Vortices with quartic potential the variation of \(m=1\) (line), \(m=1.5\) (dashed), and \(m=2\) (dots).}
    \label{fig:MH-single-quartic}
\end{figure}

For the case of generalized MH vortex where \(\kappa=0\), we have a similar profile of condensate density, \(|\phi|^2\), and the magnetic field, \(B\), shown in Figure \ref{fig:MH-single} for the sextic potential case and Figure \ref{fig:MH-single-quartic} for the quartic potential case. Here, we assume the relation between \(h\) and \(w\) mentioned in \eqref{relatehwMH} and plot the profiles with variation in \(m\). We can see that the difference between CSH and MH vortex is that the MH vortex does not emit any radial electric fields. Such property is well-known for the vortex solutions within the Ginzburg-Landau Theory of superconductivity and it turns out that such property does not have to be assumed, but instead, is an implication of the BPS property of the vortex solution. The comparative properties of the quartic and sextic potential cases with the coupling constant satisfying \eqref{relatehwMH} are also similar to the CSH vortices.

\subsubsection{generalized MCSH vortex}
\begin{figure}[h!]
    \centering
    \includegraphics[width=0.32\textwidth]{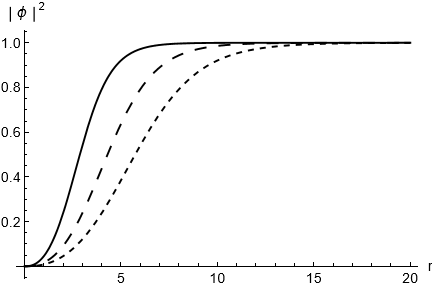}
    \includegraphics[width=0.32\textwidth]{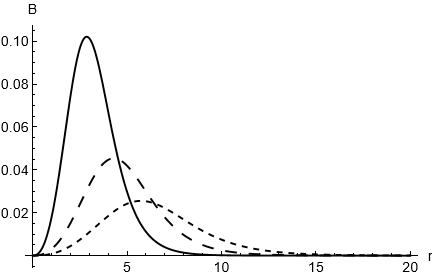}
    \includegraphics[width=0.32\textwidth]{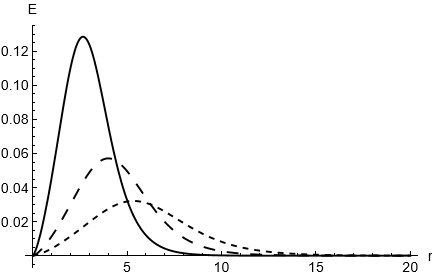}
    \caption{The profile electric fields, magnetic fields and condensate density of MCSH Vortices with sextic potential and variation of \(\kappa=1\) (line), \(\kappa=1.5\) (dashed), and \(\kappa=2\) (dots).}
    \label{fig:MCSH-single-vark}
\end{figure}

\begin{figure}[h!]
    \centering
    \includegraphics[width=0.3\textwidth]{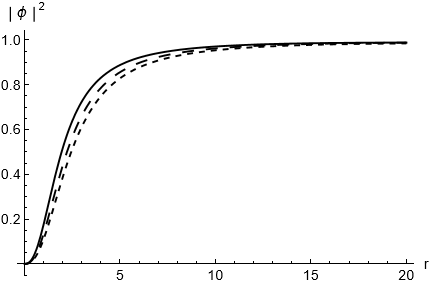}
    \includegraphics[width=0.3\textwidth]{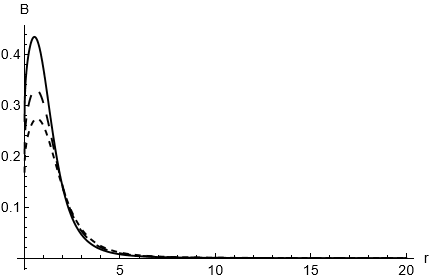}
    \includegraphics[width=0.3\textwidth]{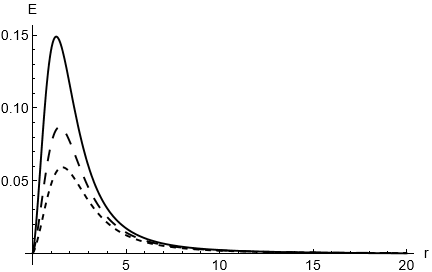}
    \caption{The profile electric fields, magnetic fields and condensate density of MCSH Vortices with quartic potential and variation of \(\kappa=1\) (line), \(\kappa=1.5\) (dashed), and \(\kappa=2\) (dots).}
    \label{fig:MCSH-single-vark-quartic}
\end{figure}

\begin{figure}[h!]
    \centering
    \includegraphics[width=0.32\textwidth]{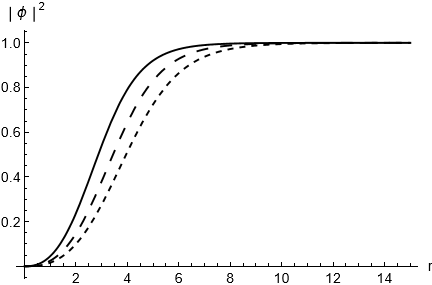}
    \includegraphics[width=0.32\textwidth]{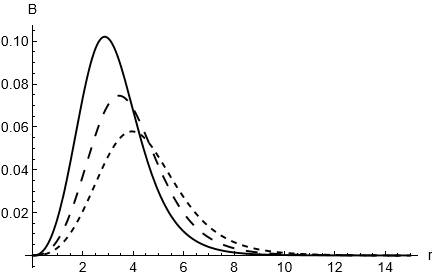}
    \includegraphics[width=0.32\textwidth]{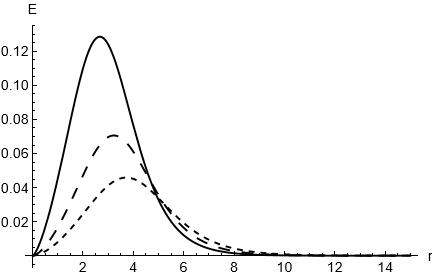}
    \caption{The profile electric fields, magnetic fields and condensate density of MCSH Vortices with sextic potential and variation of \(l=1\) (line), \(l=2\) (dashed), and \(l=3\) (dots).}
    \label{fig:MCSH-single-varl}
\end{figure}

\begin{figure}[h!]
    \centering
    \includegraphics[width=0.32\textwidth]{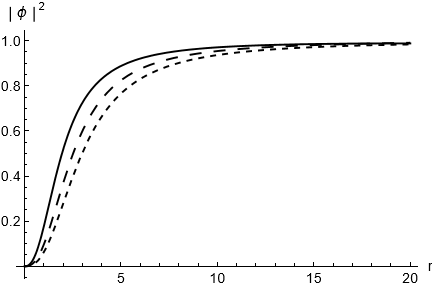}
    \includegraphics[width=0.32\textwidth]{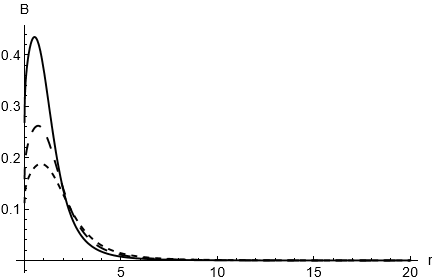}
    \includegraphics[width=0.32\textwidth]{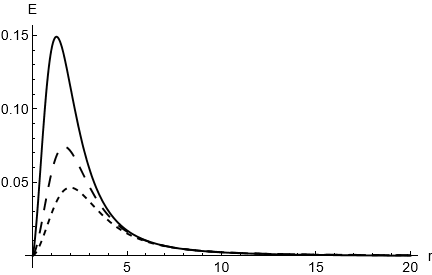}
    \caption{The profile electric fields, magnetic fields and condensate density of MCSH Vortices with quartic potential and variation of \(l=1\) (line), \(l=2\) (dashed), and \(l=3\) (dots).}
    \label{fig:MCSH-single-varl-quartic}
\end{figure}

For the generalized MCSH vortex, we have to free parameters that can be tuned in order to match the profile of the physical vortex. This implies that the model has more degree of freedom and can be effectively used to model the profile of a single vortex in a wide range of superconducting materials. In Figure \ref{fig:MCSH-single-vark} we show the profile of the density \(|\phi|\) and the electromagnetic fields for the case of sextic potential and Figure \ref{fig:MCSH-single-vark-quartic} for the quartic materials with variations of \(\kappa\). We also provide the plots with variations of \(l\) for both sextic and quartic potentials in Figure \ref{fig:MCSH-single-varl} and Figure \ref{fig:MCSH-single-varl-quartic}, respectively.

We can see that, the addition of the CS term in the Lagrangian brings back the radial electric field to this vortex, although the profiles of the fields are similar with both CSH and MH vortices. Thus, we can conclude that the role of the CS term in the generalized MCSH model is to produce the non-zero radial electric field for a vortex. This electric field, in general, can generate different interaction behaviours in a system of interacting vortices. Such a property is interesting in the context of the vortex as a topological soliton because, even though the CS term does not contribute to the total energy of the system, it provides a different dynamical picture from the usual vortex modelled by only MH terms in the Lagrangian.


\subsection{Interacting Two Vortex Solutions}
\begin{figure}[h!]
    \centering
    \includegraphics[width=0.35\textwidth]{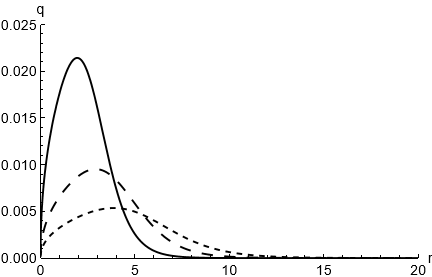}
    \includegraphics[width=0.35\textwidth]{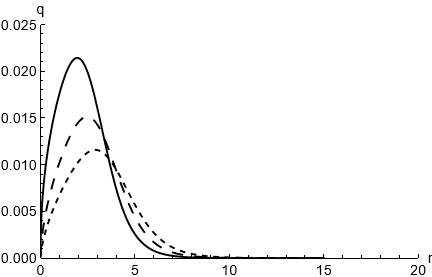}
    \caption{Topological charge density with variations of \(\kappa\) (left) and variations of \(l\) (right).}
    \label{fig:TopDens}
\end{figure}

As shown in the profile of electromagnetic fields and the condensate density in the previous section for all the BPS vortices considered, the effects of the vortices on their surroundings are similar. Thus for the numerical computation of interacting vortices, we are only going to consider the MCSH vortices, for simplicity. In order to examine the interaction between two vortices, we must first study the separability of two vortices. This can be done by looking at the profile of topological charge density which represents the "shape" of the vortex in space. The topological charge density of the vortices is given by 
\begin{equation}
    q=\frac{i}{2\pi}\epsilon^{0ij}\partial_i\phi\partial_j\bar{\phi}=\frac{n}{\pi}\frac{gg'}{r}.
\end{equation}
The profile of the topological charge density is shown in Figure \ref{fig:TopDens} where the first figure is the profile for MCSH vortex for \(\kappa\in{1,1.5,2}\), and the second figure is the profile of MCSH vortex with \(l=1,2,3\). 
Here, we can see that the topological charge density is effectively zero for \(r>10\) and beyond this radius, the value of the topological charge density is less than \(1\%\) of its maximum. Thus, we can consider two vortices separated by a distance \(d>20\) to be well separated. Such well-separated vortices interact only in long range and the non-linearity which came from the solitonic property of the vortices can be neglected. From this assumption of well-separation, we conclude that the electromagnetic fields, \(B\) and \(\textbf{E}\), from the two vortices can be superposed, and the profile of the condensate density near the centre of each vortex is given by the solution of a condensate density of a single vortex with total condensate density satisfies the Abrikosov product \cite{ABRIKOSOV1957199,Manton:2004tk},
\begin{equation}
    \phi=\prod_i\phi_i,~~~i\in\{1,2,\dots,N\},
\end{equation}
\(N\) is the number of vortices, in order to keep the asymptotic boundary condition as the stable vacuum \(|\phi|^2=1\).
As such, we can now consider the vortices as point objects interacting through their electromagnetic fields. We assume that this well-separation assumption works up to \(d=10\) where the small "tails" of the topological charge density of two vortices overlap but the values of the density are still close to the vacuum expectation value where we expect that interesting phenomena can be observed in this region. We are going to consider unit topological number vortices, \(n=1\), and \(\kappa=l=1\) from this point in order to simplify the numerical study, without any loss of generality.

\begin{figure}[h]
    \centering
    \includegraphics[width=0.4\textwidth]{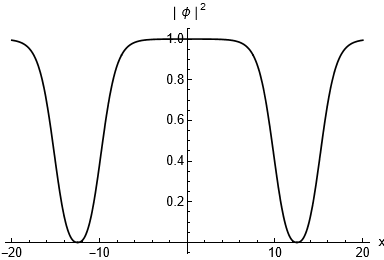}
    \includegraphics[width=0.4\textwidth]{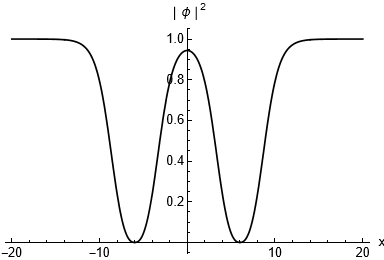}
    \includegraphics[width=0.4\textwidth]{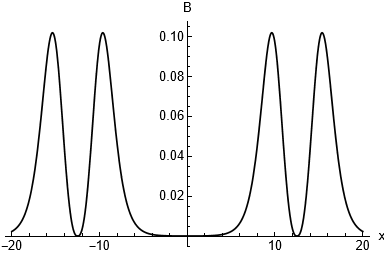}
    \includegraphics[width=0.4\textwidth]{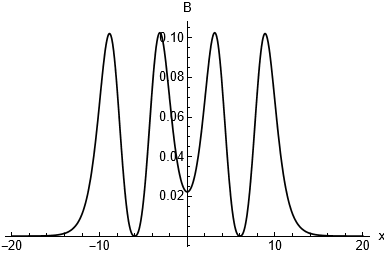}
    \caption{Plot of \(|\phi|^2\) and \(B\) for well-separated vortices (left) and close vortices (right) with \(\kappa=l=n=1\).}
    \label{fig:TwoVortices}
\end{figure}

For example, here is the plot for the density \(|\phi|^2\), and the magnetic field \(B\) for two vortices, given in Figure \ref{fig:TwoVortices}, where we assume that the separation is colinear with the horizontal axes.

The interaction property of this system with two vortices can be studied from the total energy as a function separation, \(d\). In order to do so, we must first calculate the energy density of the system by using the assumption of well-separation, such that the total condensate density satisfies, \(|\phi|^2=|\phi_1\phi_2|^2\), the total electric field satisfies superposition principle \(\textbf{E}=\textbf{E}_1+\textbf{E}_2\), and the total magnetic field satisfies \(B=B_1+B_2\), where the set \((\phi_1,B_1,\textbf{E}_1)\) are the fields from the first vortex located at \((-d/2,0)\) and \((\phi_2,B_2,\textbf{E}_2)\) are the fields of the second vortex located at \((d/2,0)\). Each of \((\phi_i,B_i,\textbf{E}_i)\) is the solution of single vortex equations. The total energy density as a function of Cartesian coordinates, \((x,y)\), written in terms of the total fields is given by
\begin{eqnarray}
    \rho_{\text{vortices}}&=&\frac{h}{2}\left(\textbf{E}\cdot\textbf{E}+B^2\right)+\frac{3l+2}{2l+1}V(|\phi|^2)+w\left(\mathcal{R}\left[\nabla\Bar{\phi}\cdot\nabla\phi\right]+\left(\frac{a^2(r_1)}{r_1^2}+\frac{a^2(r_2)}{r_2^2}\right)|\phi|^2\right),\nonumber\\
\end{eqnarray}
where \(r_1^2=(x-(d/2))^2+y^2\), \(r_2^2=(x+(d/2))^2+y^2\), \(a(r)\) is the solution of a single-vortex equation, and we have again assumed the well-separated system of vortices and the ansatz \eqref{ansatz} such that \(\Bar{\phi}A^\mu\partial_\mu\phi\approx in|\phi|^2A^\theta\), \(\textbf{A}_1\cdot \textbf{A}_2\approx0\) everywhere, and the relation between \(f\) and \(V\) is satisfied everywhere. The derivative operator, \(\nabla\), is defined as \(\nabla=\hat{\textbf{x}}\partial_x+\hat{\textbf{y}}\partial_y\).

\begin{figure}[h]
    \centering
    \includegraphics[width=0.4\textwidth]{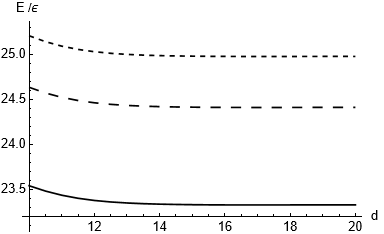}
    \includegraphics[width=0.4\textwidth]{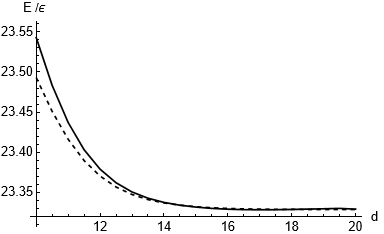}
    \caption{(\textbf{Left}) The total energy of two-vortex system with \(\kappa=n=1\) and variations  of \(l=1\) (line), \(l=2\) (dashed), and \(l=3\) (dots). (\textbf{Right})  The total energy of two-vortex systems with \(\kappa=l=1\) for the vortex-vortex system (line) and vortex-anti-vortex (dashed).}
    \label{fig:totEnerg}
\end{figure}
The total energy (in units of the characteristic energy \(\mathcal{E}\)) of two vortices with unit topological charge, \(n=1\), and \(l\in\{1,2,3\}\) is plotted against the separation distance, \(d\), for the sextic potential case of MCSH vortices, shown in Figure \ref{fig:totEnerg}. From the numerical solution above in Figure \ref{fig:totEnerg}, we can see that the total energy is asymptotic to a constant value for large \(d\). As expected from the BPS property of the vortices, it does not undergo any long-range interaction and when they are well-separated, the total energy of the system saturates the BPS bound, i.e. the total energy is linear to the total topological charge. We can also see that the Abrikosov product ansatz starts to fail at \(d<16\) but the errors of using this semi-linearized approach are still very small compared to the BPS energy, which in this case, the errors are less than \(1\%\). Thus, we can conclude that the non-linearity of the vortex as a soliton starts to dominate at distances less than \(r=8\) from the centre of the vortex. Furthermore, From the comparison between vortex-vortex and vortex-anti-vortex in the second figure of Figure \ref{fig:totEnerg}, we can conclude that both configurations obey the BPS property with the Abrikosov ansatz fails faster for vortex-vortex case that has a steeper energy increase when \(d\) becomes smaller. This is due to the fact that, although the dominant contribution to the total energy came from the scalar terms in the energy density, the electromagnetic interaction for the vortex-anti-vortex configuration is negative, hence the vortex-anti-vortex configuration has smaller errors compared to the vortex-vortex configuration within the well-separation assumptions. We can use this result to deduce that the useful summation of the topological charge of multi-vortex systems follows
\begin{eqnarray}
    |n_{\text{tot}}|=\sum_i |n_i|,
\end{eqnarray}
where \(n_i\) is the topological charge of the \(i\)-th vortex, such that the total energy of well-separated vortices satisfies \(E\propto |n_{\text{tot}}|\).

The lack of interaction between the vortices above implies that they possess particle-like properties. This can be used to model the dynamics of collective vortices as a two-dimensional ideal gas system as long as the number density of the vortex is not too high. Furthermore, because of the stable nature of vortices, lattice structures of vortices can be realized. It is interesting to study such systems because their energy behaves like an ideal gas system where the total energy is approximately linear to the number of vortices and they are a good model for predicting the distribution of electromagnetic fields on a wide range of two-dimensional superconductors because they have two free parameters which can be tuned uniquely. As such, the study of vortex lattices in the generalized MCSH model, especially as a model for superconductors, will be addressed in future works.
\section{Conclusions and Outlook}
In section III we have explained the details of the BPS Lagrangian method that is proposed to be the unified approach to produce all BPS vortex solutions in the \(U(1)\) gauge-scalar theory that is described by the generalized Maxwell-Chern-Simons-Higgs model. It is shown that from the sets of Bogomol'nyi equations and constraint equations from the BPS Lagrangian method, we can reproduce all known vortex solutions in section IV. The known solutions can be classified by their free parameters that are related to the auxiliary functions in the BPS Lagrangian, \((X_0,X_1,X_2,X_3)\), and each parameter came from the coupling of the submodels within the generalized MCSH model, such as the Chern-Simons-Higgs model and the Maxwell-Higgs model.

Such a unified approach implies the existence of a map that connects all vortex solutions in the \(U(1)\) gauge-scalar theory on the level of the dynamical field equations, in contrast with what is believed before that all of them are connected only on the Lagrangian level. This provides us with a new way to classify the topologically supported solitonic solutions within the gauge-scalar theory, especially in \(1+2\) spacetime dimensions. It can be seen in Table \ref{tab:equations} that the field equations of the known vortices follow a single pattern, even though the degree of freedom in their corresponding models are different, for example in the CSH model where we only have one free parameter, compared to the MCSH model where we can tune two free parameters. Thus, it is interesting to study how we can construct the map that connects those vortices through \(X_i\)s in more general settings where ansatz and symmetries are not assumed. Such approaches have been done with the BPS Lagrangian method for the more general non-spherically symmetric settings \cite{Fadhilla:2020rig,Atmaja:2020iti,Atmaja:2023ing}, hence we shall expect that this more generalized treatment can be used to find the vortex solutions in non-spherically symmetric generalized MCSH model. This is going to be addressed elsewhere.

We also studied the properties of the vortices from each model through numerical simulations for both single vortex settings and interacting two vortex settings in section V. We find that all of the vortices considered within this spherically symmetric ansatz possess the same profile of fields and these profiles respond to the variations of their corresponding free parameters. For example, the CSH profiles become broader in space for higher values of coupling constant \(\kappa\). As such, the MCSH vortex is the most suited known solution to model the physical vortex, for example in a two-dimensional superconductor, because it has two free parameters which can be tuned uniquely for a wide range of systems. For the interacting vortex settings, we find that two vortices do not experience any long-range and they behave like free particles when their separation is more than a certain characteristic distance which depends on their free parameters. As such, a system of vortices with low number density can be considered as an ideal gas with total energy linear to the total topological charge which came from the BPS bound saturation. Another possible configuration for a system of vortices is the vortex lattice, which can be found when the density of the vortex is high enough that the vortices stabilize themselves into a periodic structure. The study of interacting vortices might find useful applications, especially for Chern-Simons superconductors that were proposed as an alternative to the standard Bardeen–Cooper–Schrieffer superconductors \cite{Randjbar-Daemi:1989ett,Antillon:1997xr,Asorey:2001wu,Horvathy:2008hd,Banerjee:2013maa,Wang:2020zwt}. This kind of superconductor is known to be useful in modelling the properties of two dimensional materials, for example, the fractional quantum Hall effect \cite{vonKlitzing:1980pdk,Tsui:1982yy,Jackiw:1984ji}. The BPS CSH vortices found in this work can be used to approximate the configuration with the lowest energy in Chern-Simons superconductors. Such application not only can be found for the already-established Ginzburg-Landau or Chern-Simons superconductors but also for more exotic superconducting states which should be modelled by the generalized MCSH model.

One possible way to extend the study of interacting vortices is by calculating the total energy for smaller distances where well-separation assumptions cannot be used. In this regime, the non-linearity of vortices starts to dominate and the Abrikosov product ansatz fails. Such a more realistic approach is useful especially to study high energetic vortex scattering and dense vortex systems. To do this, at least two problems need to be solved, the first one is to study how to deal with the non-linear nature of vortex since it belongs to the family of topological solitons, and the second one is to study how such interaction could alter the topological charge or other conservation properties. Again, we are going to address such problems in future works.

\acknowledgments
The work in this paper is supported by GTA Research Group ITB. 
B. E. G. acknowledges the financial support through BRIN Visiting Researcher Programme. E. S. F. also would like to acknowledge the support from BRIN through the Research Assistant Programme 2023.
\appendix
\section{Second-Order Field Equations of The Model}\label{sec:SecondOrderEq}
From Section \ref{sec:MCSH} we know that the ansatz for the Higgs field and the gauge fields are given by
\begin{equation}
	\phi(r,\theta) =  g(r)e^{in\theta};\qquad \textbf{A}(r) = -\frac{\hat{\theta}}{er}\big(a(r)-n\big);\qquad A_{t} = f(r).
\end{equation}
From this ansatz, the non-trivial component of equation (\ref{Gauge-dynamics}) can be written as
\begin{align}
	&\label{Gauss-law}h\left(\frac{d^2 f}{d r^2} + \frac{1}{r}\frac{d f}{d r}\right) + \frac{d h}{d r}\frac{df}{dr} - 2{e^2}{g^2}w f  + \frac{\kappa}{e r}\frac{da}{dr} = 0,\\
	&\label{Ampere-law}h\left(\frac{d^2 a}{dr^2} - \frac{1}{r}\frac{da}{dr}\right) + \frac{dh}{dr}\frac{da}{dr} - 2{e^2}{g^2}wa + \kappa e r\frac{df}{dr} = 0.
\end{align}
Besides, equation (\ref{Higgs-dynamics}) can be written as
\begin{align}\label{higgs-profile}
	&w\left(\frac{{d^2}g}{d {r^2}} + \frac{1}{r}\frac{dg}{dr} - \frac{{a^2}g}{r^2} + {e^2}{f^2}g\right) + \frac{1}{4}\frac{dh}{dg}\left(\left(\frac{df}{dr}\right)^{2} - \left(\frac{1}{er}\frac{da}{dr}\right)^{2}\right)\nonumber\\
	&\qquad\qquad\qquad + \frac{1}{2}\frac{dw}{dg}\left(\left(\frac{dg}{dr}\right)^{2} - \frac{{a^2}{g^2}}{r^2} + {e^2}{f^2}{g^2}\right) - \frac{1}{2}\frac{dV}{dg} = 0
\end{align}
where we have $w\equiv w(g)$, $h\equiv h(g)$, and $V \equiv V(g)$. All of the Bogomol'nyi equations satisfy these second-order equations.
\section{Derivation of The Constraint Equations For the MH Vortex}
In this section, we demonstrate the details of how we can derive the constraints for \(X_i\)s in the MH vortex via the BPS Lagrangian method. We can produce the equations for MH vortex by taking \(\kappa=0\), such that the zero discriminant constraint becomes
\begin{subequations}
\begin{align}
    {r^{-2}} &: a^2 g^2 w - \frac{X_1^2}{4w} + \frac{X_3^2}{2h} = 0,\label{eq:ConstraintEqM-1}\\
    {r^0} &: V -  f^2 g^2 w - X_0 - \frac{ X_2^2}{2h} = 0. \label{eq:ConstraintEqM-2}
\end{align}
\end{subequations}
and the constraints from Euler-Lagrange equations of \(\mathcal{L}_{BPS}\) become
\begin{subequations}
\begin{align}
    &\frac{\partial X_0}{\partial f} = \frac{X_2}{h}\left(\frac{\partial X_3}{\partial a} - \frac{\partial X_2}{\partial f}\right),\label{eq:EulerLagrange-M-1}\\
    &0 = \frac{X_1}{2w}\left(\frac{\partial X_3}{\partial g} - \frac{\partial X_1}{\partial f}\right),\label{eq:EulerLagrange-M-2}\\
    &0 = \frac{\partial X_0}{\partial a},\label{eq:EulerLagrange-M-3}\\
    &\frac{X_1}{2w}\left(\frac{\partial X_2}{\partial g} - \frac{\partial X_1}{\partial a}\right) = \frac{X_3}{h}\left(\frac{\partial X_2}{\partial f} - \frac{\partial X_3}{\partial a}\right),\label{eq:EulerLagrange-M-4}\\
    &\frac{\partial X_0}{\partial g} = \frac{ X_2}{h}\left(\frac{\partial X_1}{\partial a} - \frac{\partial X_2}{\partial g}\right),\label{eq:EulerLagrange-M-5}\\
    &0 = \frac{X_3}{h}\left(\frac{\partial X_1}{\partial f} - \frac{\partial X_3}{\partial g}\right).\label{eq:EulerLagrange-M-6}
\end{align}
\end{subequations}

From equation (\ref{eq:EulerLagrange-M-3}), we find that $X_0 \equiv X_0 (g,f)$. Consequently, from (\ref{eq:ConstraintEqM-2}) we know $X_2 \equiv X_2 (g,f)$. The fact that $X_0$ does not depend on $a(r)$ may also be used for equation (\ref{eq:EulerLagrange-M-1}) and (\ref{eq:EulerLagrange-M-4}) that we may write \(X_1\) nad \(X_3\) as \eqref{X1-Y} and \eqref{X3-Z}. These may be substituted to equation (\ref{eq:ConstraintEqM-1}) so that \begin{equation}
    a^2 \left(g^2 w - \frac{Y_1^2}{4w} + \frac{Z_1^2}{2h}\right) + a \left(\frac{Y_1 Y_2}{2w} + \frac{Z_1Z_2}{h}\right) - \frac{Y_2^2}{4w} + \frac{Z_2^2}{2h} = 0 .
\end{equation}
Since all the coefficients of the above equation do not depend on $a(r)$, we have
\begin{subequations}
\begin{align}
    g^2 w - \frac{Y_1^2}{4w} + \frac{Z_1^2}{2h} &= 0,\\
    \frac{Y_1 Y_2}{2w} + \frac{Z_1Z_2}{h} &= 0,\label{eq:ConstraintEqM-1-a1}\\
    - \frac{Y_2^2}{4w} + \frac{Z_2^2}{2h} &= 0 .\label{eq:ConstraintEqM-1-a0}
\end{align}
\end{subequations}
We may also take the first derivative for $f(r)$ to equation (\ref{eq:ConstraintEqM-2}). This gives
\begin{equation}
    -2f g^2 w - \frac{\partial X_0}{\partial f} - \frac{2 X_2}{2h}\frac{\partial X_2}{\partial f} = 0.
\end{equation}
Substituting equation (\ref{eq:EulerLagrange-M-1}) to the above result gives
\begin{equation}\label{eq:X2-X3Tilde}
    X_2 Z_1 = -2 f g^2 w h.
\end{equation}
Writing $Z_1$ in term of $X_2$, we may use (\ref{eq:ConstraintEqM-1-a2}) to write $Y_1$ as below.
\begin{equation}\label{eq:X1Tildesquared}
    Y_1^2 = 4g^2 w^2 + \frac{8 f^2 g^4 w^3 h}{X_2^2} .
\end{equation}
The above result is then substituted into (\ref{eq:ConstraintEqM-1-a1}) to obtain
\begin{equation}
    Y_2^2 \left(4 g^2 w^2 + \frac{8 f^2 g^4 w^3 h}{X_2^2}\right) = Z_2^2 \frac{16 f^2 g^4 w^4}{X_2^2} .
\end{equation}
With (\ref{eq:ConstraintEqM-1-a0}), the result obtained above translated into
\begin{equation}
    \frac{2 g^2 w^2}{h}Z_2^2 = 0 ,
\end{equation}
from which it implies $Z_2 = 0$ and consequently $Y_2 = 0$.

Now, consider the equation \eqref{eq:EulerLagrange-M-2} which, under consideration of above results, translates to 
\begin{eqnarray}
    \frac{\partial Z_1}{\partial g}=\frac{\partial Y_1}{\partial f}.
\end{eqnarray}
From the perspective of the equations \eqref{eq:X2-X3Tilde} and \eqref{eq:X1Tildesquared}
we have
\begin{eqnarray}
    \frac{\partial Z_1}{\partial g}&=&-\frac{2f}{X_2}\frac{\partial(g^2 w h)}{\partial g}+\frac{2fg^2wh}{X^2_2}\frac{\partial X_2}{\partial g},\\
    \frac{\partial Y_1}{\partial f}&=&\frac{8fg^4w^3h}{Y_1 X_2}\frac{\partial}{\partial f}\left(\frac{f}{X_2}\right),
\end{eqnarray}
hence
\begin{equation}
    -2f \left(\frac{\partial}{\partial g}\left(\frac{g^2 w h}{X_2}\right) + \frac{4g^4 w^3 h}{Y_1 X_2}\frac{\partial}{\partial f}\left(\frac{f}{X_2}\right)\right) = 0 .
\end{equation}
In terms of \(Z_1\) and \(Y_1\), this equation can be recast into
\begin{equation}
    f \left(\frac{\partial}{\partial g}\left(\frac{Z_1}{f}\right) - \frac{2g^2 w^2 Z_1}{f Y_1}\frac{\partial}{\partial f}\left(\frac{Z_1}{g^2wh}\right)\right) = 0 .
\end{equation}
This result gives two possible cases.

\section{Derivation of The Constraint Equations For The MCSH Vortex}
In this subsection, we demonstrate that the BPS Lagrangian method is able to reproduce the known MCSH vortex solution proposed in \cite{Andrade:2021qkq}. By substituting \eqref{X1-Y} and \eqref{X3-Z} to \eqref{const-eq3} and \eqref{const-eq5} we arrive at the following equations,
\begin{align}
	&\frac{\partial X_0}{\partial g}(g,f) = \frac{2{X_2} - \kappa f}{2h}\left({Y_1}(g,f) - \frac{\partial X_2}{\partial g}\right)\label{DgX0}\\
	&\frac{\partial X_0}{\partial f}(g,f) = \frac{2{X_2} - \kappa f}{2h}\left({Z_1}(g,f) - \frac{\partial X_2}{\partial f}\right)\label{DfX0}
\end{align}
Through (\ref{const-eq2}) and (\ref{DfX0}), one may obtain
\begin{equation}\label{X2}
	X_2 = \frac{\kappa f}{2} + \frac{4f{g^2}wh}{\kappa - 2{Z_1}}
\end{equation}
and
\begin{equation}\label{X0-V}
	{X_0} = V - \frac{8{f^2}{g^4}{w^2}h}{(\kappa - 2{Z_1})^2} - {f^2}{g^2}w
\end{equation}

Besides, by the requirement equation of positive definite condition of the generalized functions $w(g)$ and $h(g)$, (\ref{const-eq1}) tells us that $X_1$ cannot be zero. Therefore (\ref{const-eq7}) implies
\begin{equation}\label{X1-X3}
	\frac{\partial X_3}{\partial g} = \frac{\partial X_1}{\partial f}
\end{equation}
which also satisfy (\ref{const-eq6}). By virtue of (\ref{X1-Y}) and (\ref{X3-Z}), the above equation can be rewritten as
\begin{equation}
	a\left(\frac{\partial Z_1}{\partial g} - \frac{\partial Y_1}{\partial f}\right) = \frac{\partial Y_2}{\partial f} - \frac{\partial Z_2}{\partial g}
\end{equation}
since the right-hand side does not depend on $a(r)$, this result implies
\begin{align}
	&\label{Z1-Y1}\frac{\partial Z_1}{\partial g} = \frac{\partial Y_1}{\partial f}\\
	&\label{Z2-Y2}\frac{\partial Z_2}{\partial g} = \frac{\partial Y_2}{\partial f}
\end{align}
One may also find that (\ref{const-eq1}) can be rewritten in terms of ${Y_1}, {Y_2}, {Z_1}$ and $Z_2$ and it can be regarded as a quadratic equation in $a(r)$. Since the coefficient does not depend on $a$ explicitly, in order for the equation to be satisfied for any $a$, we must have
\begin{align}
	&\frac{Y_1^2}{4w} - \frac{\kappa^2}{8h} - \frac{\kappa Z_1}{2h} - \frac{Z_1^2}{2h} - {g^2}w = 0\\
	&\frac{n{\kappa^2}}{4h} + \frac{{Y_1}{Y_2}}{2w} + \frac{\kappa n {Z_1}}{2h} - \frac{\kappa {Z_2}}{2h} - \frac{{Z_1}{Z_2}}{h} = 0\label{const-eq10}\\
	&\frac{\kappa n {Z_2}}{2h} + \frac{Y_2^2}{4w} - \frac{{\kappa^2}{n^2}}{8h} - \frac{Z_2^2}{2h} = 0\label{const-eq11}
\end{align}

The remaining non-trivial constraint equations are (\ref{const-eq3}) and (\ref{const-eq8}). We may substitute (\ref{X1-Y}) and (\ref{X3-Z}) into these two equations. Upon substitution, by the same argument that we used above, we obtain two constraint equations that are
\begin{align}\label{const-a^1}
	\frac{Y_1}{2w}\frac{\partial}{\partial g}\left(\frac{\kappa f}{2} + \frac{4f{g^2}wh}{\kappa - 2{Z_1}}\right) - \frac{Y_1^2}{2w} - \frac{2Z_1+\kappa}{2h}\left(\frac{\kappa}{2} + \frac{4{g^2}wh}{\kappa - 2{Z_1}} - {Z_1}\right)= 0
\end{align}
and
\begin{align}\label{const-a^0}
	\frac{2Z_2-\kappa n}{2h}\left(\frac{\kappa}{2} + \frac{4{g^2}wh}{\kappa - 2{Z_1}} - {Z_1}\right) + \frac{{Y_1}{Y_2}}{2w} - \frac{Y_2}{2w}\frac{\partial}{\partial g}\left(\frac{\kappa f}{2} + \frac{4f{g^2}wh}{\kappa - 2{Z_1}}\right) = 0.
\end{align}
We may write the above equations as
\begin{equation}
	\left(\frac{{Z_1}{Y_2}}{h} + \frac{\kappa Y_2}{2h} - \frac{{Y_1}{Z_2}}{h} + \frac{\kappa n {Y_1}}{2h}\right)\left(\frac{\kappa}{2} + \frac{4{g^2}wh}{\kappa - 2{Z_1}} - {Z_1}\right) = 0
\end{equation}
Since the second multiplicative term of the above expression does not give real constraint function $Z_1$, the only possibility left is
\begin{equation}\label{const-eq12}
	\frac{{Z_1}{Y_2}}{h} + \frac{\kappa Y_2}{2h} - \frac{{Y_1}{Z_2}}{h} + \frac{\kappa n {Y_1}}{2h} = 0
\end{equation}
Solving the above equation for $Z_2$ and then substitute it into (\ref{const-eq10}), we have
\begin{equation}
	-\frac{{Y_2}\left(w(\kappa + 2{Z_1})^{2} - 2h{Y_1^2}\right)}{4hw{Y_1}} = 0
\end{equation}
which gives two possibilities, that is
\begin{equation}
	2h{Y_1^2} = w\left(\kappa + 2{Z_1}\right)^{2} \qquad \text{or}\qquad {Y_2} = 0
\end{equation}
One may check that the first possibility does not satisfy the rest of the constraint. This left us with the second possibility. Upon substitution to (\ref{const-eq11}), it gives $Z_2 = \frac{n\kappa}{2}$. This result satisfy equation (\ref{const-eq10}), (\ref{const-eq11}), (\ref{const-a^0}), and (\ref{const-eq12}).

\medskip
$\bibliographystyle{utphys}$
\bibliography{MCSH}

\end{document}